\pdfoutput=1
\documentclass{jfm}
\usepackage{graphicx}
\usepackage{epstopdf, epsfig}
\usepackage[]{subcaption} 
\usepackage{color}
\usepackage{placeins} 
\usepackage{amsmath} 
\usepackage{natbib}
\usepackage[colorlinks=true,allcolors=blue]{hyperref}

\shorttitle{Near-wake structure of full-scale vertical-axis wind turbines}
\shortauthor{N. J. Wei, I. D. Brownstein, J. L. Cardona, M. F. Howland, and J. O. Dabiri}

\title{Near-wake structure of full-scale vertical-axis wind turbines}

\author{Nathaniel J. Wei\aff{1},
  Ian D. Brownstein\aff{2},
  Jennifer L. Cardona\aff{1},
  Michael F. Howland\aff{1},
   \and John O. Dabiri\aff{3}
   \corresp{\email{jodabiri@caltech.edu}}}

\affiliation{\aff{1}Department of Mechanical Engineering, Stanford University, Stanford, California 94305, USA
\aff{2}XFlow Energy Company, Seattle, Washington 98108, USA
\aff{3}Graduate Aerospace Laboratories \& Mechanical Engineering, California Institute of Technology, Pasadena, California 91125, USA}

\begin{document}

\maketitle

\begin{abstract}

To design and optimize arrays of vertical-axis wind turbines (VAWTs) for maximal power density and minimal wake losses, a careful consideration of the inherently three-dimensional structure of the wakes of these turbines in real operating conditions is needed. Accordingly, a {new} volumetric particle-tracking velocimetry method was developed to measure three-dimensional flow fields around full-scale VAWTs in field conditions. Experiments were conducted at the Field Laboratory for Optimized Wind Energy (FLOWE) in Lancaster, CA, using six cameras and artificial snow as tracer particles. Velocity and vorticity measurements were obtained for a 2-kW turbine with five straight blades and a 1-kW turbine with three helical blades, each at two distinct tip-speed ratios and at Reynolds numbers based on the rotor diameter $D$ between $1.26 \times 10^6$ and $1.81 \times 10^6$. A tilted wake was observed to be induced by the helical-bladed turbine. By considering the dynamics of vortex lines shed from the rotating blades, the tilted wake was connected to the geometry of the helical blades. Furthermore, the effects of the tilted wake on a streamwise horseshoe vortex induced by the rotation of the turbine were quantified. Lastly, the implications of these dynamics for the recovery of the wake were examined. This study thus establishes a fluid-mechanical connection between the geometric features of a VAWT and the salient three-dimensional flow characteristics of its near-wake region, which can {potentially} inform both the design of turbines and the arrangement of turbines into highly efficient arrays.

\end{abstract}


\section{Introduction}

Wind turbines are becoming increasingly important contributors to global energy supplies, as they represent a low-carbon alternative to traditional power-generation technologies that rely on the combustion of fossil fuels. If wind power is to comprise a larger share of global energy production, the efficiency and power density of wind farms will need to be improved \citep{jacobson_saturation_2012}. The critical limitation of these large arrays is not the efficiency of individual wind turbines, which already operate at efficiencies approaching their theoretical maximum \citep{betz_maximum_1920}, but rather the dynamics of wind-turbine wakes and their effects on downstream turbines \citep{stevens_flow_2017}. The efficiency of the large-scale deployment of wind power thus depends largely on a more careful consideration of the wake dynamics of wind turbines.

Horizontal-axis wind turbines (HAWTs) typically have very long wakes that extend up to 20 turbine diameters ($D$) downstream of the turbine itself \citep{vermeer_wind_2003,meyers_optimal_2012,hau_wind_2013,stevens_flow_2017}, {though wake persistence may vary with field conditions \cite[e.g.][]{nygaard_wake_2018}}. HAWTs placed in closely packed arrays therefore generally incur significant losses from the wakes of upstream turbines \cite[e.g.][]{barthelmie_modelling_2007,barthelmie_evaluation_2010,barthelmie_quantifying_2010}. Vertical-axis wind turbine (VAWT) wakes, by contrast, have been observed to recover their kinetic energy within 4 to 6 $D$ downstream of the turbine, albeit with lower individual coefficients of thrust and power \citep{kinzel_energy_2012,kinzel_turbulence_2015,ryan_three-dimensional_2016}. VAWTs can also be arranged in pairs, to capitalize on synergistic fluid interactions between the turbines \cite[e.g.][]{rajagopalan_aerodynamic_1990,brownstein_performance_2016,ahmadi-baloutaki_wind_2016,hezaveh_increasing_2018,brownstein_aerodynamically_2019}. Taken together, these factors imply that arrays of VAWTs can potentially achieve power densities an order of magnitude higher than those of conventional wind farms \citep{dabiri_potential_2011}.

The dynamics of VAWT wakes are therefore relevant to the design of large-scale wind farms with higher energy densities, and accordingly have been analyzed in several recent studies. The replenishment of momentum in both HAWT and VAWT wakes has been shown to be dependent on turbulent entrainment of fluid from above the turbine array, through modeling \citep{meneveau_top-down_2012,luzzatto-fegiz_entrainment_2018}, simulations \citep{calaf_large_2010,hezaveh_mean_2018}, and field experiments \citep{kinzel_turbulence_2015}. VAWTs also exhibit large-scale vortical structures in their wakes that may further augment wake recovery. These vortex dynamics have been observed in scale-model studies of varying geometric fidelity, from rotating circular cylinders \citep{craig_kinematic_2016} to complete rotors \cite[e.g.][]{tescione_near_2014,brownstein_aerodynamically_2019}. The inherently three-dimensional nature of these vortical structures, coupled with the high Reynolds numbers of operational VAWTs, complicates experimental and numerical studies of the dynamics of VAWT wakes.

Accordingly, numerical simulations with varying levels of complexity have been applied to study VAWT wakes. For studies of wake interactions within arrays, 2D Reynolds-averaged Navier-Stokes (RANS) simulations have often been employed \cite[e.g.][]{bremseth_computational_2016,zanforlin_fluid_2016}. Large-eddy simulation (LES) studies have generally used actuator-line models to approximate the effects of the individual blades on the flow \cite[e.g.][]{shamsoddin_large_2014,shamsoddin_large-eddy_2016,abkar_self-similarity_2017,hezaveh_mean_2018,abkar_impact_2018}. \cite{posa_wake_2016} and \cite{posa_large_2018} were able to resolve the unsteady vortex shedding of individual blades in the spanwise component of vorticity using LES with periodic boundary conditions in the spanwise direction, which meant that tip-vortex shedding was not captured. {More recently, \cite{villeneuve_improving_2020} used delayed detached-eddy simulations (DDES) and a fully three-dimensional turbine model in a rotating overset mesh to study the effects of end plates on VAWT wakes, resolving 3D vortex shedding and wake dynamics up to 10 $D$ into the wake.} Numerical simulations have thus {continued to improve in their} capacity to resolve the salient dynamics in the wakes of VAWTs. 

Experimental studies of wake structures of VAWTs have generally been limited to planar measurements in the laboratory \cite[e.g.][]{brochier_water_1986,ferreira_visualization_2009}. The deployment of stereoscopic particle-image velocimetry (stereo-PIV) has allowed some three-dimensional effects to be captured \citep{tescione_near_2014,rolin_wind-tunnel_2015}, as has the use of planar PIV with multiple imaging planes \citep{parker_effect_2016,parker_effect_2017,araya_transition_2017}. These planar techniques have been successful in identifying characteristic vortex phenomena, such as dynamic stall on turbine blades \citep{ferreira_visualization_2009,dunne_dynamic_2015,buchner_dynamic_2015,buchner_dynamic_2018} and tip-vortex shedding from the ends of individual blades \citep{hofemann_3d_2008,tescione_near_2014}. Generally, however, it is difficult to compute all three components of vorticity or the circulation of vortical structures with purely planar measurements. The analysis of the three-dimensional character of vortical structures in the wake is therefore greatly facilitated by fully three-dimensional flow-field measurements. Such experiments have only recently been carried out in laboratory settings. Using tomographic PIV, \cite{caridi_hfsb-seeding_2016} resolved the three-dimensional structure of tip vortices shed by a VAWT blade within a small measurement volume with a maximum dimension of 5.5 blade chord lengths. \cite{ryan_three-dimensional_2016} obtained 3D time-averaged velocity and vorticity measurements of the full wake of a model VAWT using magnetic-resonance velocimetry (MRV). Most recently, \cite{brownstein_aerodynamically_2019} used 3D particle-tracking velocimetry (PTV) to obtain time-averaged measurements of the wakes of isolated and paired VAWTs. These studies were all carried out at laboratory-scale Reynolds numbers, which fell between one and two orders of magnitude below those typical of operational VAWTs. \cite{miller_vertical-axis_2018} attained Reynolds numbers up to $Re_D = 5\times10^6$ using a compressed-air wind tunnel, and their measurements of the coefficients of power demonstrated a Reynolds-number invariance for $Re_D > 1.5\times10^6$. These power measurements suggest that Reynolds numbers on the order of $10^6$ may be required to fully capture the wake dynamics of field-scale turbines. As an alternative to laboratory experiments at lower $Re_D$, experiments in field conditions at full scale are possible, but these have been limited to pointwise anemometry \citep{kinzel_energy_2012,kinzel_turbulence_2015} or 2D planar velocity measurements \citep{hong_natural_2014}. Thus, {for the validation and extension of} existing experimental work on the wake dynamics of VAWTs, 3D flow measurements around full-scale VAWTs in field conditions are desirable.

The additional benefit of 3D flow measurements is that they enable the effects of complex turbine geometries on wake structures to be investigated. This is particularly useful for VAWTs, since several distinct geometric variations exist. Because of the planar constraints of most experimental and numerical studies, VAWTs with straight blades and constant spanwise cross-sections have primarily been studied due to their symmetry and simplicity of construction. However, there exist several VAWT designs that incorporate curved blades. Large-scale Darrieus-type turbines have blades that are bowed outward along the span, {and these have historically reached larger sizes and power-generation capacities than straight-bladed turbines \citep{mollerstrom_historical_2019}}. Similarly, many modern VAWTs have helical blades that twist around the axis of rotation, following the design of the Gorlov Helical Turbine \citep{gorlov_united_1995}, to reduce fatigue from unsteady loads on the turbine blades. Relatively few studies have investigated the flow physics of these helical-bladed VAWTs in any kind of detail. \cite{schuerich_effect_2011} used a vorticity-transport model for this purpose, and \cite{cheng_aerodynamic_2017} approached the problem using unsteady RANS, 2D LES, and wind-tunnel experiments. Most recently, \cite{ouro_three-dimensionality_2019} analyzed turbulence quantities in the wake of a helical-bladed VAWT in a water channel using pointwise velocity measurements from an anemometer on a traverse. A full treatment of the effects of the helical blades on the three-dimensional vorticity fields and corresponding vortical structures in the wake has yet to be undertaken.

Thus, the purpose of this work is to study the three-dimensional flow features of operational VAWTs in the field. {The work revolves around two primary contributions: the development of a field-deployed 3D-PTV measurement system, and the analysis of topological and dynamical characteristics of vortical structures in the wakes of full-scale VAWTs}. {First}, the characterisation of {the artificial-snow based technique} for obtaining three-dimensional, three-component measurements of velocity and vorticity in field experiments around full-scale VAWTs will be documented (section \ref{sec:sec2}). Experiments with two VAWTs, one with straight blades and one with helical blades and each at two tip-speed ratios, will {then} be outlined. Velocity and vorticity fields from these experiments will be presented, and a difference in the three-dimensional structure of the wake {between the two types of turbines} will be identified (sections \ref{sec:sec3_velocities} and \ref{sec:sec3_vorticities}). This tilted-wake behavior will be analyzed further, to establish a connection between blade geometry and wake topology (section \ref{sec:sec3_twist}). The results from these analyses shed light on the dynamics that govern the near wake of VAWTs {(up to $\sim3$ $D$ downstream, cf.\ \cite{araya_transition_2017})}, and implications of these findings for future studies and for the design and optimization of VAWT wind farms will be discussed (sections \ref{sec:sec3_dynamics} and \ref{sec:sec3_implications}). This work represents the first full-scale, fully three-dimensional flow-field study on operational VAWTs in field conditions, and therefore provides fundamental insights on wake dynamics at high Reynolds numbers and the fluid mechanics of wind energy.


\section{Experimental Methods}
\label{sec:sec2}

In this section, the setup of the field experiments is outlined, and a {novel technique for} 3D-PTV measurements {in field conditions is introduced. The experimental procedure is discussed, and} the post-processing steps for computing {velocity and} vorticity fields are described. {Additional details and characterisations of the measurement system are given in appendices \ref{sec:appendix_snow} and \ref{sec:appendix_processing}.}

\subsection{Field Site and Turbine Characterization}
\label{sec:sec2_field}

\begin{figure}
\begin{subfigure}[t]{0.48\textwidth}
\centering
  \includegraphics[width=\textwidth]{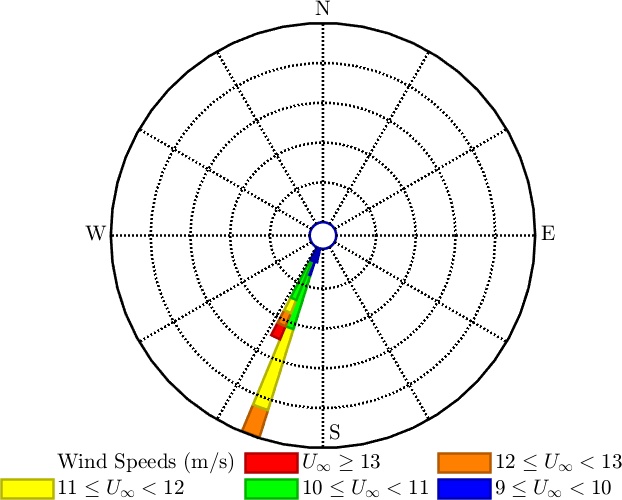}
  \caption{}
\label{fig:windRose}
\end{subfigure}
\hfill
\begin{subfigure}[t]{0.48\textwidth}
\centering
  \includegraphics[width=\textwidth]{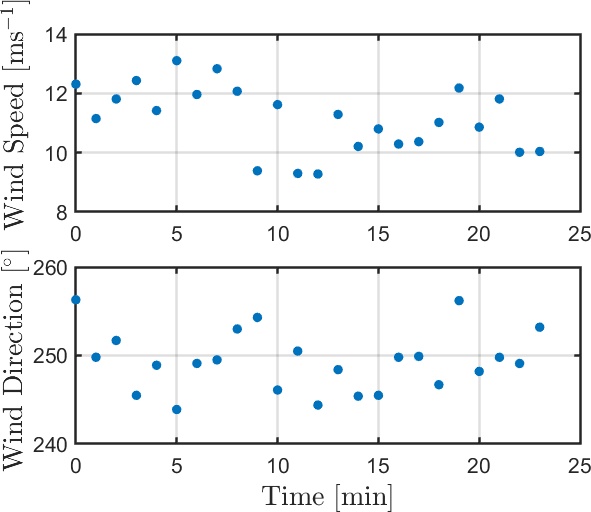}
  \caption{}
\label{fig:wind1min}
\end{subfigure}
\caption{(a) Wind rose for conditions during experiments (9-11 August 2018). The plotted wind speeds and directions are those recorded by the tower-mounted anemometer, located 10 m above the ground, {and have been binned} in ten-minute averages by 1 $\rm{ms^{-1}}$ and $5^\circ$, respectively. {(b) Wind speeds and directions, from a single experiment (three eight-minute data sets), binned in one-minute averages.}}
\label{fig:windStats}
\end{figure}

Experiments were carried out during the nights of 9-11 August 2018 at the Field Laboratory for Optimized Wind Energy (FLOWE), located on a flat, arid segment of land near Lancaster, California, USA. Details regarding the geography of the site are provided by \cite{kinzel_energy_2012}. Wind conditions at the site were measured with an anemometer (First Class, Thies Clima) and a wind vane (Model 024A, Met One), which recorded data at 1 Hz with accuracies of $\pm 3\%$ and $\pm 5^\circ$, respectively. These were mounted on a meteorological tower (Model M-10M, Aluma Tower Co.) at a height of 10 m above the ground. The tower also recorded air temperature, which was used to interpolate air density from a density-temperature table. A datalogger (CR1000, Campbell Scientific) recorded these data at 1- and 10-minute intervals. The height difference between the anemometer and the VAWTs was corrected using a fit of an atmospheric boundary-layer profile to data collected previously at the site at multiple heights \cite[cf.][]{kinzel_energy_2012}. The correction resulted in a 3\% change in the free-stream velocity, which compared more favorably with the particle-based flow-field measurements than the uncorrected readings. The measured wind conditions at the site during experiments were uniform in both magnitude and direction: the wind speed was $11.01\pm1.36$ $\rm{ms^{-1}}$, and the wind direction was from the southwest at $248\pm 3^\circ$. These statistics were calculated from sensor data that had been averaged by the datalogger into ten-minute readouts, and are summarized in the wind rose shown in figure \ref{fig:windRose}.

\begin{figure}
	\begin{minipage}{0.4\textwidth}
		\centering
		\includegraphics[width=0.45\textwidth]{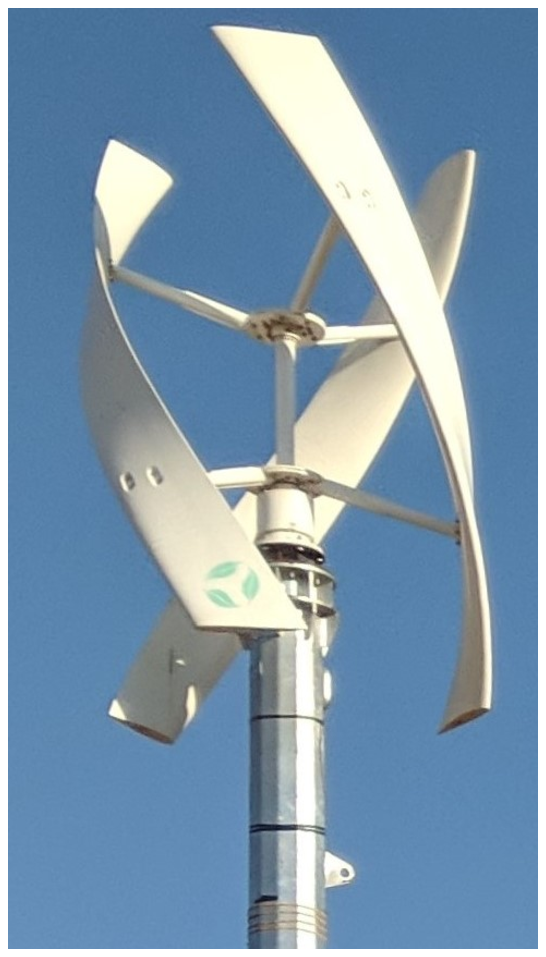}
	\end{minipage}\hfill
	\begin{minipage}{0.6\textwidth}
		\centering
		Helical-Bladed Turbine (UGE)
		\begin{tabular}{l|l}
		\hline
		    Manufacturer & Urban Green Energy \\
			Model & VisionAir 3 \\
			Rated Power & 1 kW \\
			Number of Blades & 3 \\
			Turbine Radius & 0.9 m \\
			Turbine Span & 3.2 m \\
			Blade Chord & 0.511 m \\
			Solidity ($\sigma$) & $0.271\pm0.030$ \\
		\end{tabular}
	\end{minipage}
	\begin{minipage}{\textwidth}
	\begin{tabular}{l}
	\end{tabular}
	\end{minipage}\hfill
	\begin{minipage}{0.4\textwidth}
		\centering
		\includegraphics[width=0.45\textwidth]{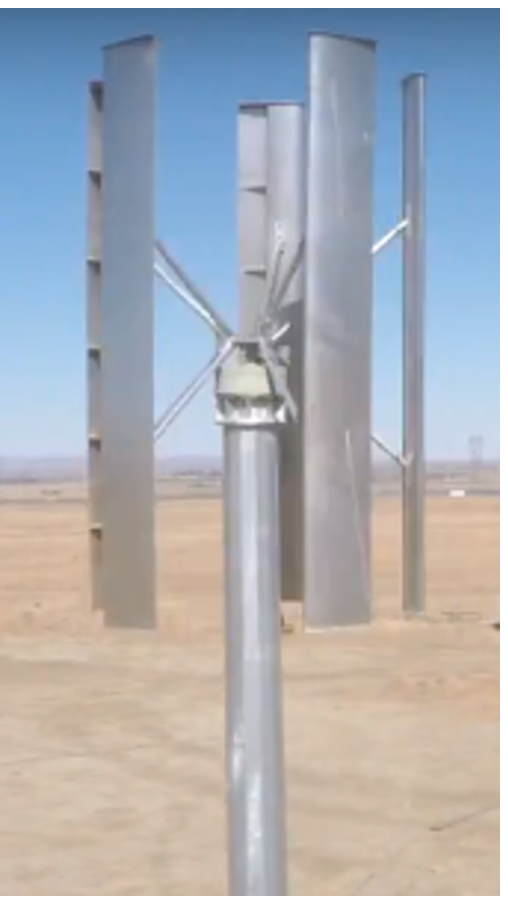}
	\end{minipage}\hfill
	\begin{minipage}{0.6\textwidth}
		\centering
		Straight-Bladed Turbine (WPE)
		\begin{tabular}{l|l}
		\hline
		    Manufacturer & Wing Power Energy \\
			Model & N/A \\
			Rated Power & 2 kW \\
			Number of Blades & 5 \\
			Turbine Radius & 1.1 m \\
			Turbine Span & 3.7 m \\
			Blade Chord & 0.483 m \\
			Solidity ($\sigma$) & $0.349\pm0.020$ \\
		\end{tabular}
	\end{minipage}
	\caption{Photographs (left) and specifications (right) of the helical-bladed UGE turbine (top) and the straight-bladed WPE turbine (bottom). The blade twist of the UGE turbine is $\tau = 0.694$ $\rm{rad\cdot m^{-1}}$.}
    \label{fig:turbines}
\end{figure}

For these experiments, two types of VAWTs were employed. A 1-kW, three-bladed VAWT with helical blades, built by Urban Green Energy (UGE), was compared with a 2-kW, five-bladed VAWT with straight blades from Wing Power Energy (WPE). The blades of both turbines had constant cross-sectional geometries. The blade twist of the helical-bladed turbine, representing the angle of twist with respect to the axis of rotation per unit length along the span of the turbine, was $\tau = 0.694$ $\rm{rad\cdot m^{-1}}$. Photos and details of these two turbines, referred to in this work by their manufacturer's acronyms (UGE and WPE), are given in figure \ref{fig:turbines}. Each turbine was tested at two different tip-speed ratios, defined as 

\begin{equation}
    \lambda = \frac{\omega R}{U_\infty},
    \label{eqn:TSR}
\end{equation}

where $\omega$ is the rotation rate of the turbine ($\rm{rad\cdot s^{-1}}$), $R$ is the radius of the turbine (m), and $U_\infty$ is the magnitude of the free-stream velocity ($\rm{ms^{-1}}$). The turbines had different solidities, quantified as the ratio of the blade area to the swept area of the rotating blades. This was defined as

\begin{equation}
    \sigma = \frac{nc}{\pi D},
\end{equation}

where $n$ is the number of blades, $c$ is the chord length of each blade, and $D$ is the turbine diameter. The parameters for the four experiments presented in this work are given in table \ref{tab:expParams}.

\begin{table}
  \begin{center}
\def~{\hphantom{0}}
  \begin{tabular}{l|cccc}
        \textbf{Case Identifier} & \includegraphics[width=12pt]{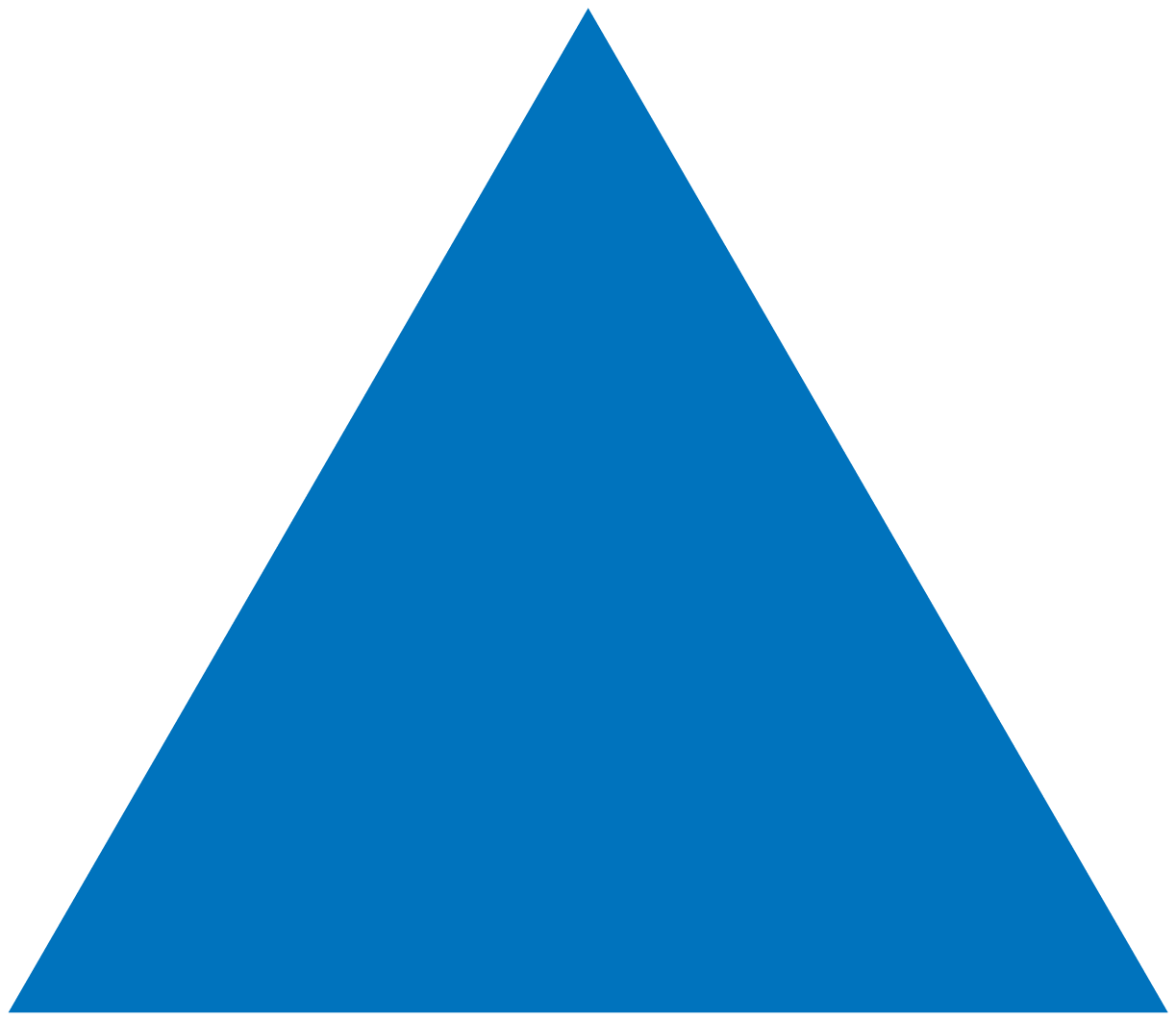} & \includegraphics[width=12pt]{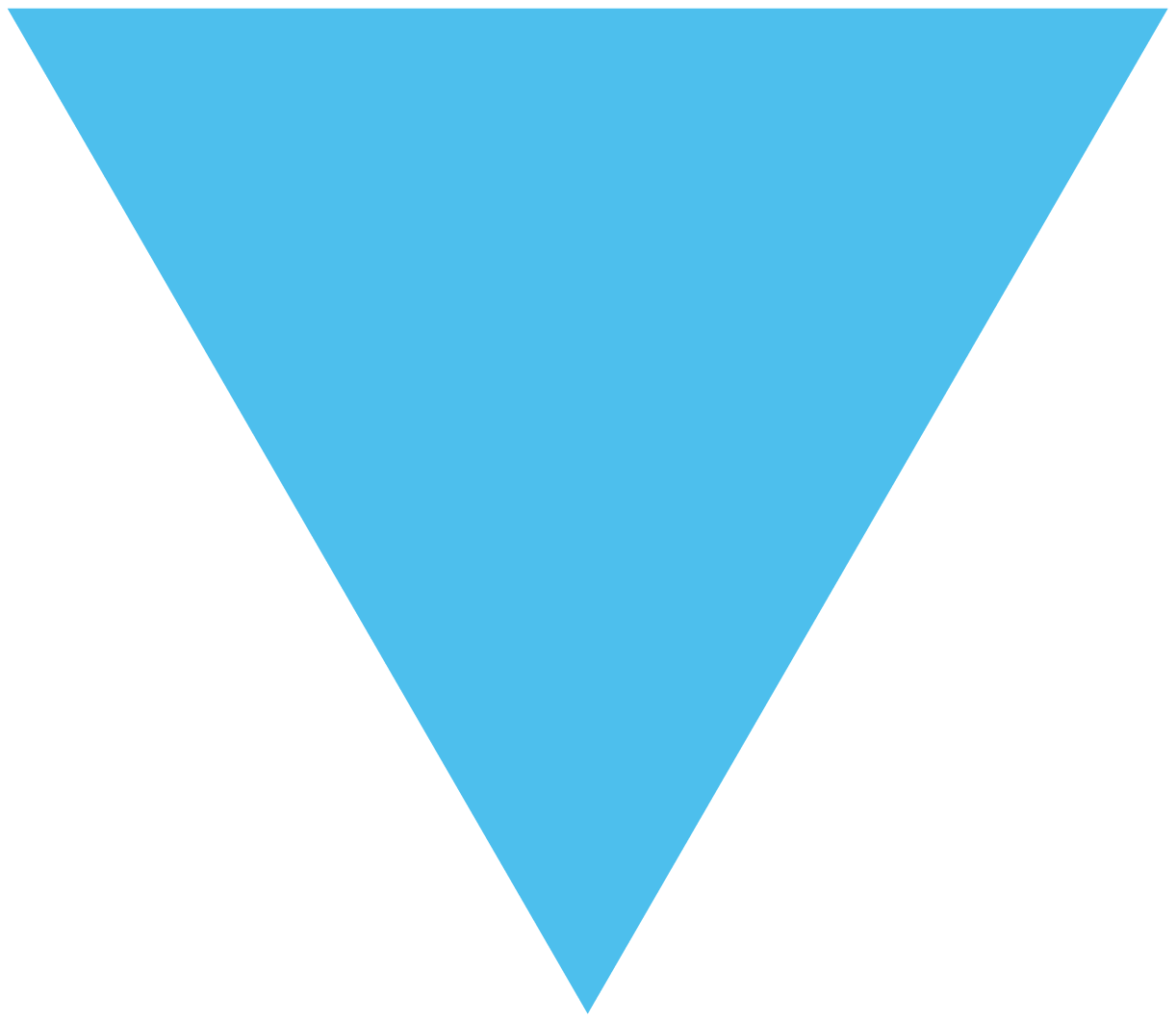} & \includegraphics[width=12pt]{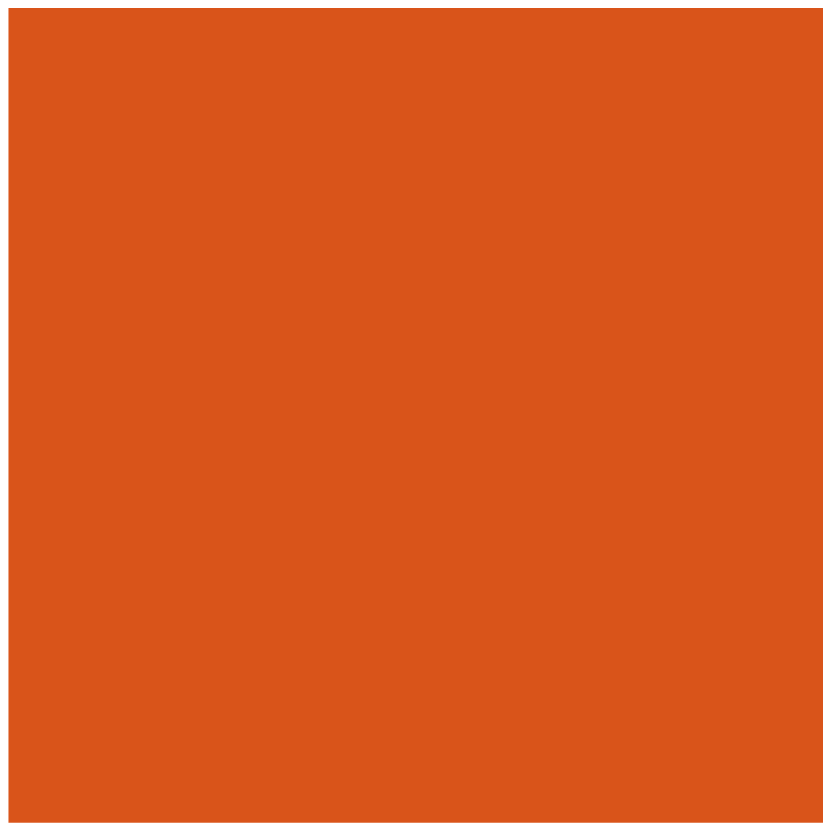} & \includegraphics[width=14pt]{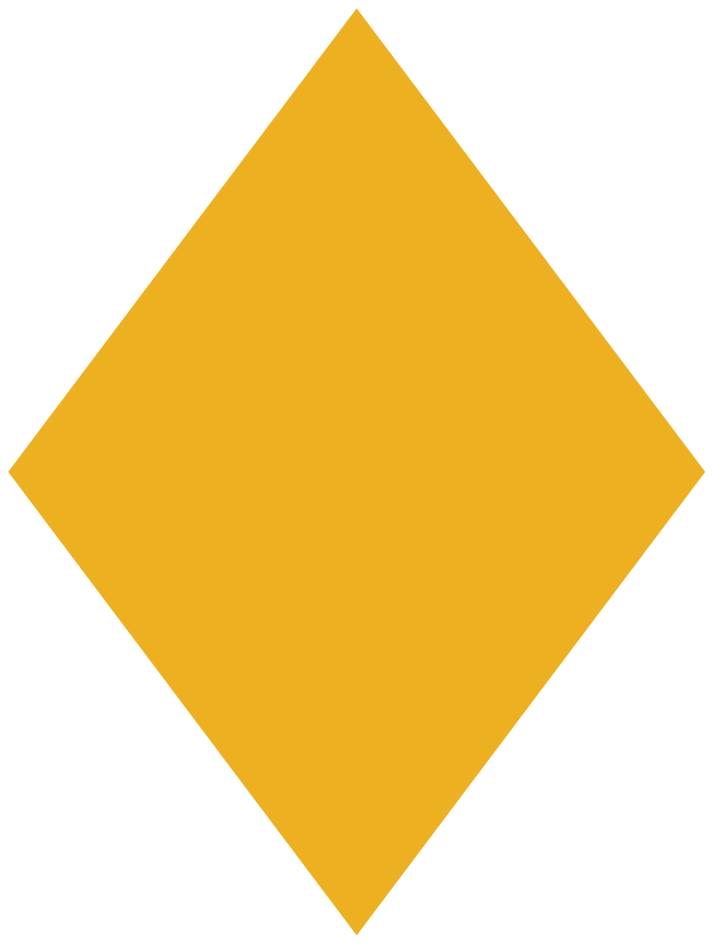} \\[6pt]
        Turbine & UGE & UGE & WPE & WPE \\
        Wind Speed, $U_\infty$ ($\rm{ms^{-1}}$) & $11.15\pm1.37$ & $11.39\pm1.25$ & $12.76\pm1.54$ & $10.10\pm1.29$ \\
        Duration, $T$ (s) & 1393 & 1343 & 1453 & 1393 \\
        Duration, $T^* = T\frac{U_\infty}{D}$ (--) & 8627 & 8491 & 8423 & 6396 \\
        Reynolds Number, $Re_D\times10^6$ & $1.26\pm0.15$ & $1.29\pm0.14$ & $1.81\pm0.22$ & $1.43\pm0.18$\\
        Tip-Speed Ratio, $\lambda$ & $1.19\pm0.13$ & $1.40\pm0.14$ & $0.96\pm0.10$ & $1.20\pm0.16$ \\
        Solidity, $\sigma$ & $0.271\pm0.030$ & $0.271\pm0.030$ & $0.349\pm0.020$ & $0.349\pm0.020$ \\
  \end{tabular}
  \caption{Experimental parameters for the four test cases presented in this work. From the left, the first two experiments were carried out on 9 August 2018, the third on 10 August, and the fourth on 11 August. The nondimensional duration $T^*$ represents the number of convective time units $D/U_\infty$ captured by each experiment. {Uncertainties from the average values represent one standard deviation over time.}}
  \label{tab:expParams}
  \end{center}
\end{table}

\begin{figure}
\centering
\includegraphics[width=0.5\textwidth]{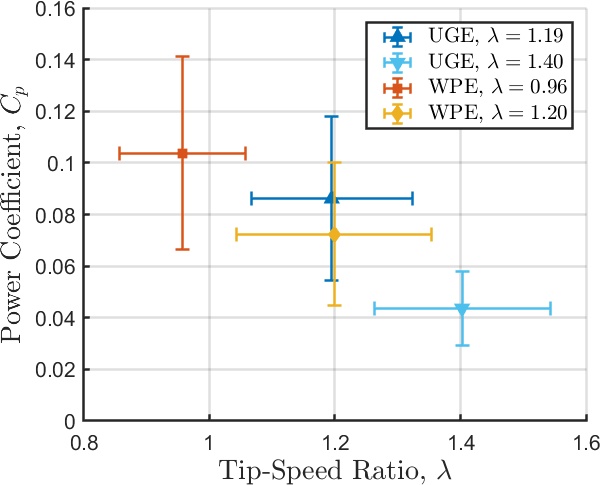}
\caption{Coefficient of power as a function of tip-speed ratio for the four experiments outlined in table \ref{tab:expParams}.}
\label{fig:Cp_TSR}
\end{figure}

The turbines were mounted on the same tower for experiments, setting the mid-span location of each at a height of 8.2 m above the ground. A Hall-effect sensor (Model 55505, Hamlin) on the tower measured the rotation rate of the WPE turbine by recording the blade passing frequency. This method could not be implemented with the UGE turbine due to its different construction. Therefore, the rotation rate of the UGE turbine was calculated from videos of the turbine in operation, taken at 120 frames per second with a CMOS camera (Hero4, GoPro), by autocorrelating the pixel-intensity signal to establish a blade passing time. Electrical power outputs from the turbines were measured and recorded in 10-minute intervals using a second datalogger (CR1000, Campbell Scientific). The coefficient of power was then calculated as

\begin{equation}
C_p = \frac{P}{\frac{1}{2}\rho SD {U_\infty}^3},
    \label{eqn:Cp}
\end{equation}

where $P$ is the power produced by the turbine, $\rho$ is the density of air, and $S$ is the turbine span. The computed coefficients of power of the two turbines for each of the tested tip-speed ratios are shown in figure \ref{fig:Cp_TSR}. The $C_p$ values for both turbines agree with measurements from previous experiments at the FLOWE field site reported by \cite{miller_vertical-axis_2018} that suggested that the optimal tip-speed ratio for maximizing $C_p$ was on the order of $\lambda \approx 1$ for the WPE turbine. This operating tip-speed ratio is low compared to those of larger-scale VAWTs with lower solidities \cite[cf.][]{mollerstrom_historical_2019}, but is consistent with those of turbines of the same power-production class \cite[e.g.][]{han_design_2018}. The results also agree with the findings of other studies that the optimal tip-speed ratio for power production decreases with increasing solidity \citep{miller_rotor_2018,rezaeiha_towards_2018}.

\subsection{Particle-Tracking Velocimetry}
\label{sec:sec2_PTV}

\begin{figure}
\begin{subfigure}[t]{0.39\textwidth}
\centering
  \includegraphics[width=\textwidth]{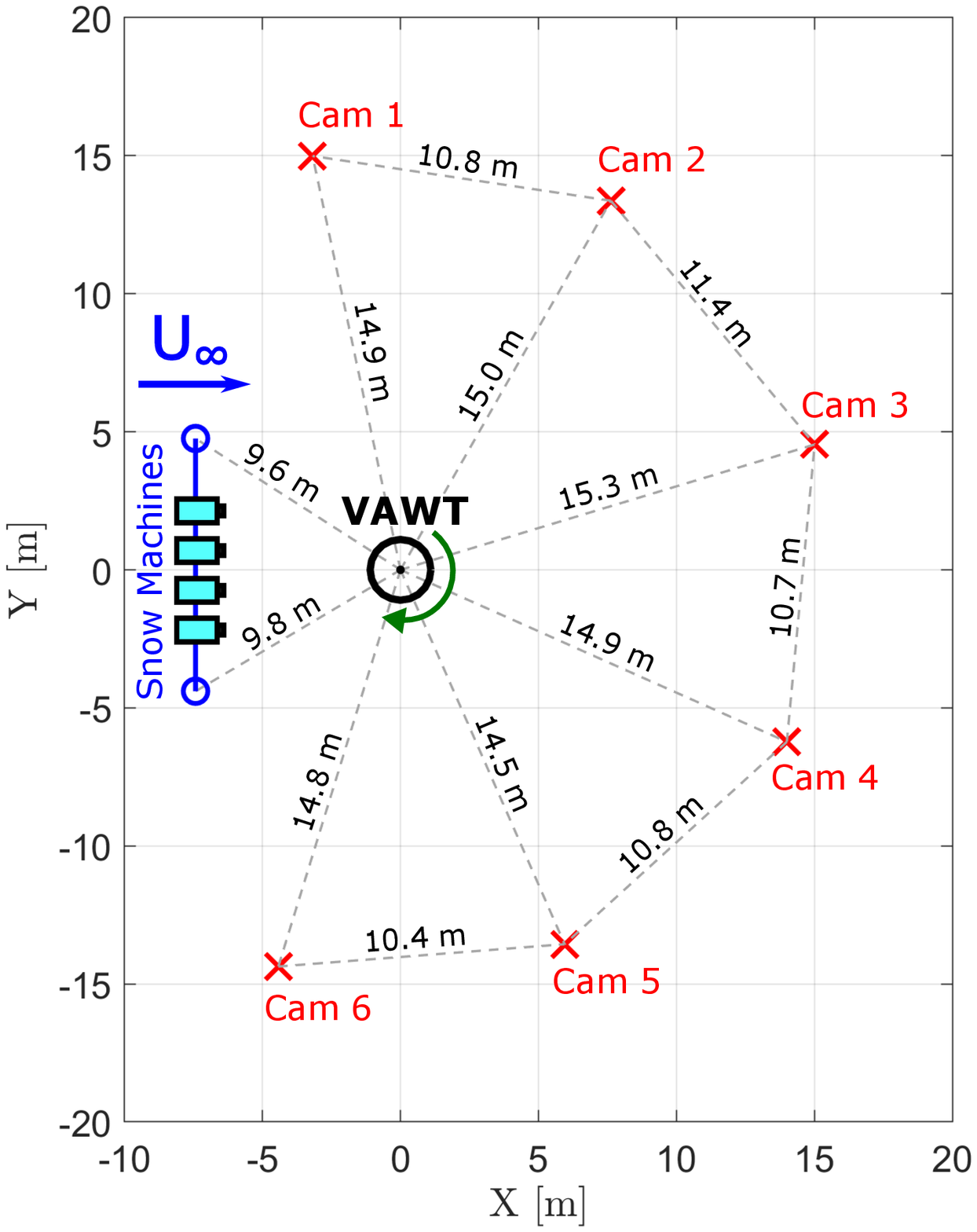}
  \caption{}
\label{fig:fieldSite2D}
\end{subfigure}
\hfill
\begin{subfigure}[t]{0.59\textwidth}
\centering
  \includegraphics[width=\textwidth]{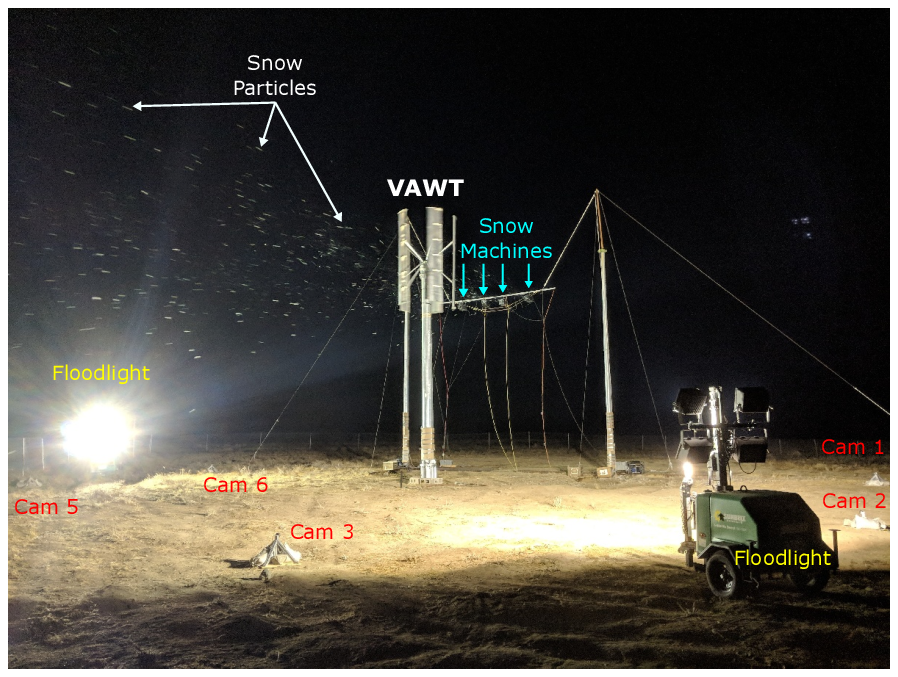}
  \caption{}
\label{fig:fieldSitePhoto}
\end{subfigure}
\caption{Schematic (a) and photograph (b) of the field experiment. {The snow machines in (a) are not drawn to scale.} The video cameras are labeled as Cam 1 through Cam 6. The direction of rotation for the turbine in the diagram is clockwise, and the Z-coordinate points vertically upward from the ground. The WPE turbine ($S = 3.7$ m) is shown in the photo. The artificial snow particles are visible moving with the flow toward the left of the frame.}
\label{fig:fieldSite}
\end{figure}

{A new method for volumetric flow measurements in field conditions was developed to obtain three-dimensional velocity measurements in a large measurement volume encapsulating the near wake of full-scale turbines. Because of the arid climate at the field site, natural precipitation could not be relied on to populate the required measurement volume with seeding particles \cite[cf.][]{hong_natural_2014}. Therefore,} to quantify the flow around the turbines, artificial snow particles were used as {seeding particles} for the flow. These were produced by four snow machines (Silent Storm {DMX, Ultratec Special Effects}) that were suspended by cables from two poles approximately four turbine diameters ($D$) upstream of the turbine tower. The machines could be raised to different heights with respect to the turbine, to adjust the distribution of particles in the measurement volume. The particles were illuminated by two construction floodlights (MLT3060, Magnum), so that their images contrasted the night sky. Six CMOS video cameras (Hero4, GoPro) were mounted on frames and installed in a semicircle on the ground to capture the particles in the turbine wake from several different angles. The layout of the entire experiment (excluding the upstream meteorological tower) is shown in figure \ref{fig:fieldSite}. The low seeding density and high visibility of the particles {under these conditions meant that unambiguous particle trajectories could be extracted from the camera images, making PTV a natural choice for obtaining three-dimensional, three-component volumetric velocity measurements.}

{The degree to which the artificial snow particles follow the flow was an important factor for the accuracy of the PTV measurements, and their aerodynamic characteristics were accordingly considered in detail. The particles were composed of an air-filled soap foam with an average effective diameter of $d_p = 11.2\pm 4.2$ mm and an average density of $\rho = 6.57\pm0.32$ $\rm{kg m^{-3}}$. Since the particles were relatively large and non-spherical, experiments were conducted in laboratory conditions to establish their aerodynamic characteristics. A detailed description of the experiments, results, and analyses regarding particle response and resulting experimental error is given in appendix \ref{sec:appendix_snow}. By releasing the particles into a wind tunnel as a jet in crossflow and tracking them using 3D-PTV with four cameras, the particle-response timescale $\tau_p$ and slip velocity $V_s$ were computed. Comparing $\tau_p$ with the relevant flow timescale, $\tau_f = D/U_\infty$, yielded a particle Stokes number of $Sk = \tau_p/\tau_f \approx 0.23$. The worst-case slip velocities for the field experiments were estimated to be $V_{s} \lesssim 0.170$ $\rm{ms^{-1}}$, or less than 2\% of the average wind speed. Therefore, the particles were found to follow the flow with sufficient accuracy to resolve the large-scale structures encountered in VAWT wakes in the field.}

To ensure that the entire measurement domain was sampled {with sufficient numbers of particles}, multiple iterations of {each experimental case} were conducted, focusing on different regions of the measurement volume. This was accomplished by raising the snow machines to three distinct heights with respect to the turbine: at the turbine mid-span, above the turbine mid-span, and at the top of the turbine (approximately 8 m, 9 m, and 10 m above the ground, respectively). Vector fields obtained from the data for each case were combined, so that each experiment listed in table \ref{tab:expParams} represents the combination of three separate recording periods. The effects of time-averaging are further discussed in section \ref{sec:sec2_analysis}.

The six video cameras were arranged in a semicircle around the near-wake region of the turbines, up to 7 to 8 $D$ downstream of the turbine. They recorded video at 120 frames per second and a resolution of $1920\times1080$ pixels, with an exposure time of 1/480 s and an image sensitivity (ISO) of 6400. The test cases outlined in table \ref{tab:expParams} thus represent between 161,000 and 175,000 images per camera of the measurement volume. Because the cameras were all positioned on the downstream side of the turbine, some flow regions adjacent to the turbine were masked by the turbine itself, and thus could not be measured. This, however, did not affect the measurement of the wake dynamics downstream of the turbine. The total measurement volume was approximately 10 m $\times$ 7 m $\times$ 7 m, extending up to 2 $D$ upstream of the turbine and at least 3 $D$ downstream into the wake.

To achieve the 3D reconstruction of {the artificial snow} particles in physical space from the 2D camera images, a wand-based calibration procedure following that of \cite{theriault_protocol_2014} was carried out. {A description of the procedure and its precision is included in appendix \ref{sec:appB_calibration}. The two calibrations collected at the field site resulted in reconstruction errors of distances between cameras of $0.74\pm0.39\%$ and $0.83\pm0.41\%$, and reconstruction errors in the spans of the turbines of $0.21\%$ and $0.32\%$. The calibrations therefore allowed particle positions to be triangulated accurately in physical space.}

\subsection{Experimental Procedure}
\label{sec:sec2_procedure}

The collection of data for the field experiments was undertaken as follows. The rotation rate of the turbine was controlled via electrical loading to change the tip-speed ratio between experiments. The snow machines were hoisted to the desired height relative to the turbine, and artificial snow particles were advected through the measurement domain by the ambient wind. All six cameras were then initiated to record 420 to 560 seconds of video. Wind speed and direction measurements from the meteorological tower were averaged and recorded in one-minute bins over the duration of the recording period. The procedure was then repeated for the three different snow-machine heights listed in section \ref{sec:sec2_PTV}, corresponding to three data sets in total for each experimental case.

\subsection{Data Processing and Analysis}
\label{sec:sec2_analysis}

{The procedure and algorithms used to obtain accurate time-averaged velocity fields from the raw camera images are presented in detail in appendix \ref{sec:appB_PTV}, along with an analysis of the statistical convergence of the averaged data. An overview of the procedure is given here.}

{Particles were isolated in the raw images through background subtraction and masking.} Because the cameras were not synchronized via their hardware, the images from the camera views were temporally aligned using an LED band (RGBW LED strip, Supernight) that was mounted on the turbine tower below the turbine blades and flashed at one-minute intervals. {Particles were then identified in the synchronized images by thresholding based on pixel intensities. The particles were mapped to 3D locations in physical space using epipolar geometry \citep{hartley_multiple_2003}. A multi-frame predictive-tracking algorithm developed by \cite{ouellette_quantitative_2006} and \cite{xu_tracking_2008} computed Lagrangian particle trajectories and velocities from these locations. Velocity data recorded with the snow machines set at different heights were combined into a single unstructured volume of instantaneous velocity vectors. This field was averaged into discrete cubic voxels with side lengths of 25 cm. The standard deviation of the velocity magnitude over all vectors in these voxels was below 5\% of the average value for over 50\% of the voxels in the measurement domain. This quantity can be interpreted as an analogue for measurement precision, though some variation due to turbulence across vectors within the voxels was expected. In the wake of the turbine, the volume of interest for the analyses presented in this work, 87\% of voxels had standard deviations below 5\%, with 58\% having values below 2\%. The best-case precision for highly sampled voxels was below 1\%. A more thorough account of these statistical-convergence studies is provided in appendix \ref{sec:appB_PTV}. The velocity fields were finally filtered to enforce a zero-divergence condition \citep{schiavazzi_matching_2014} and were used to compute vorticity fields.}

It is important to briefly consider the effect of time-averaging, which was inherent to this PTV method. Temporal averaging removed the presence of turbulence fluctuations in the data, and thus the resulting velocity and vorticity fields necessarily contained only flow phenomena that were present in the mean flow. Conjectures regarding the effects of turbulence fluctuations on momentum transfer into the wake are therefore not possible based on these measurements. {These effects are expected to dominate the far wake, whereas the near wake is characterised by large-scale vortical structures. Therefore, for the purposes of this study, it was deemed acceptable to forgo the resolution of turbulence fluctuations.} Similarly, unsteady vortex dynamics were not resolved in these experiments. However, the effects of vortex dynamics occurring at large scales can still be observed in the time-averaged flow fields, as will be shown in the following section. It is thus possible to infer connections between the time-averaged results of these experiments and the underlying unsteady dynamics observed in previous studies \cite[e.g.][]{tescione_near_2014,parker_effect_2016,ryan_three-dimensional_2016,araya_transition_2017}.

\section{Experimental Results}
\label{sec:sec3}

In this section, the results of the field experiments described in the previous section are presented and further analyzed. First, the velocity fields for the four experimental cases given in table \ref{tab:expParams} are shown to highlight three-dimensional flow features. Next, the vorticity fields for these cases are presented, and a tilted wake is observed for the helical-bladed turbine. The tilted wake is then analyzed further, and a connection between turbine-blade geometry and wake tilt is established. Finally, the dynamics of the identified vortical structures are considered, and the implications of the results for arrays of VAWTs are outlined.

\subsection{Velocity Fields}
\label{sec:sec3_velocities}

\begin{figure}
\centering
  \includegraphics[width=\textwidth]{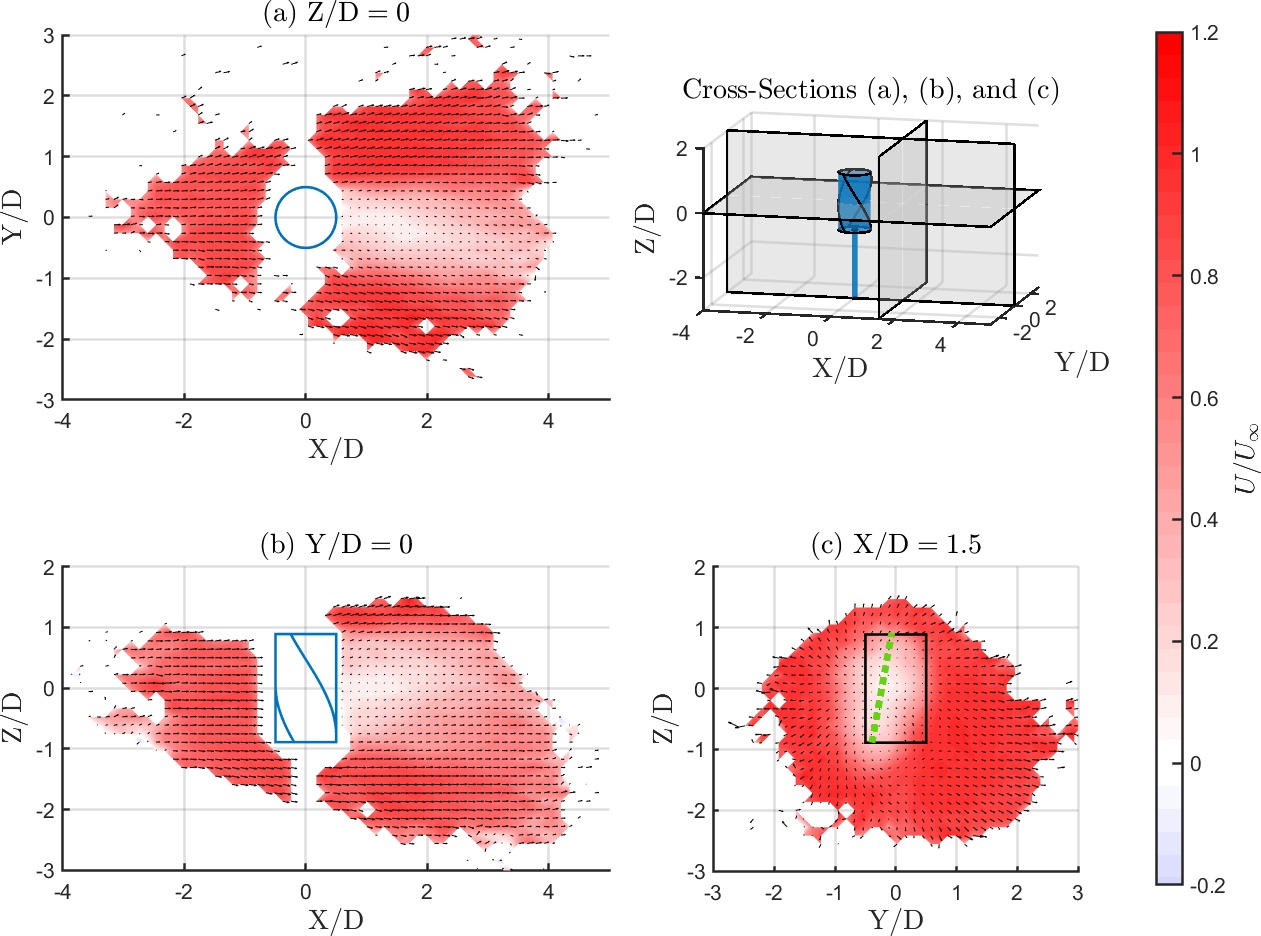}
\caption{Three orthogonal time-averaged planar fields of the streamwise velocity $U$ for the helical-bladed turbine, taken at $Z/D = 0$, $Y/D = 0$, and $X/D = 1.5$ (counter-clockwise, from top left). A slight tilt from the vertical in the clockwise direction, {shown by a fit to minima in the streamwise velocity (dashed green line),} is visible in the velocity-deficit region in the $YZ$ cross-section.}
\label{fig:u_UGE}
\end{figure}

\begin{figure}
\centering
  \includegraphics[width=\textwidth]{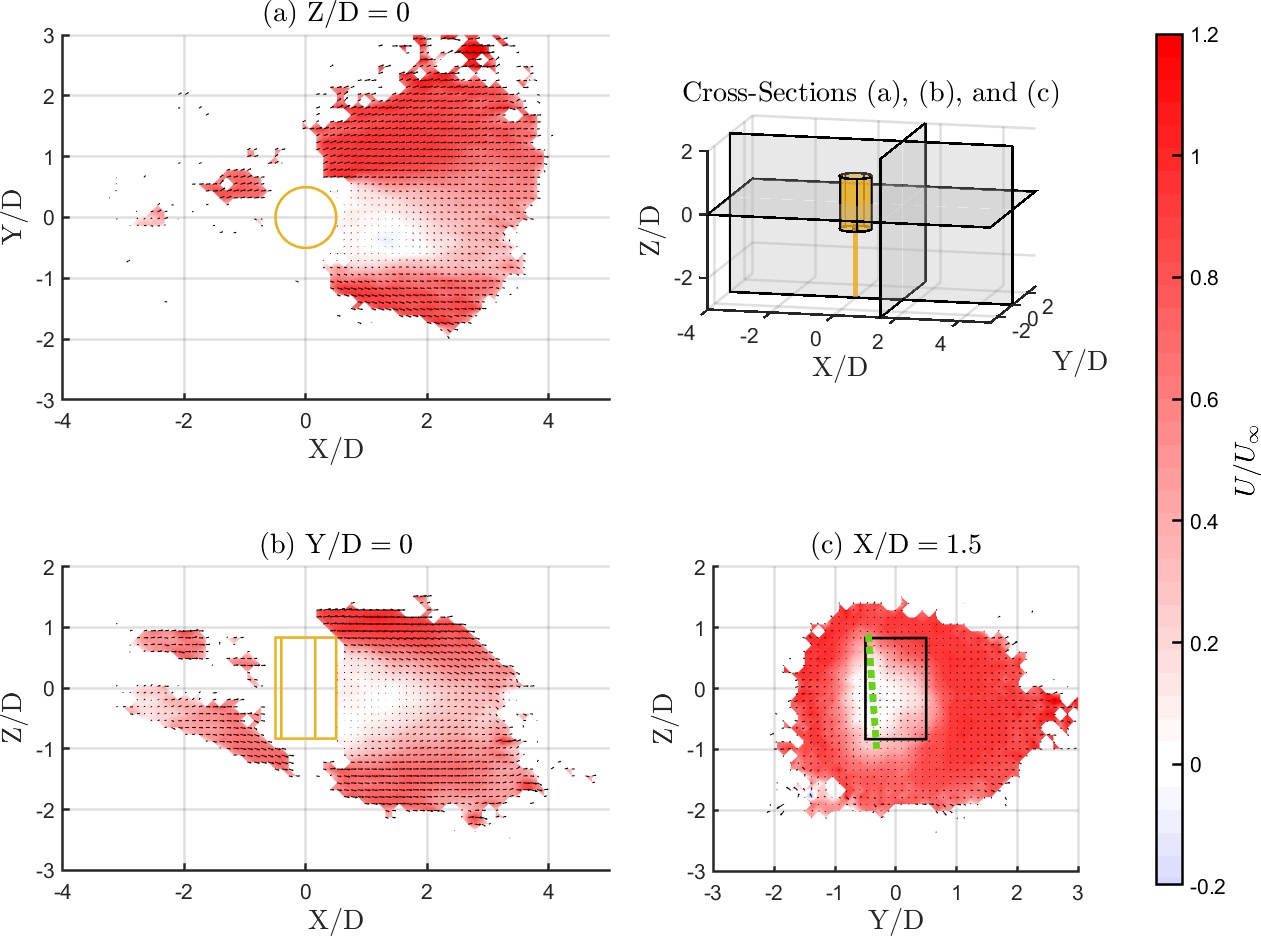}
\caption{Three orthogonal time-averaged planar fields of the streamwise velocity $U$ for the straight-bladed turbine, taken at $Z/D = 0$, $Y/D = 0$, and $X/D = 1.5$ (counter-clockwise, from top left). In contrast to figure \ref{fig:u_UGE}, no wake tilt is present in the $YZ$ cross-section, {as evidenced by the relatively vertical alignment of the fit to the wake profile (dashed green line)}.}
\label{fig:u_WPE}
\end{figure}

Velocity fields for the time-averaged streamwise-velocity component $U$ on three orthogonal planar cross-sections are given in figures \ref{fig:u_UGE} and \ref{fig:u_WPE} for the helical-bladed (UGE) and straight-bladed (WPE) turbines at tip-speed ratios of $\lambda = 1.19$ and $\lambda = 1.20$, respectively. The wake structure was not observed to change significantly with the changes in $\lambda$ achieved in these experiments. A more detailed analysis of the wake velocity fields for both turbines, including comparisons with previous wake studies at lower $Re_D$, is provided in appendix \ref{sec:appC_velocity}. As this study seeks to ascertain the effects of turbine blade geometry on the 3D structure of the near wake, two key differences between the velocity fields of the two turbines are highlighted.

First, in $YZ$ cross-sections of $U$ downstream of the turbine, topological differences were evident in the wake region, where the local streamwise velocity fell below the free-stream velocity. This region was observed to tilt in the clockwise direction when viewed from downstream in the case of the helical-bladed turbine (figure \ref{fig:u_UGE}), while no such tilt was observed in the case of the straight-bladed turbine (figure \ref{fig:u_WPE}). This tilted-wake behavior will be analyzed in detail in section \ref{sec:sec3_twist}, and it will be shown that this topological difference was a consequence of the blade shape of the helical-bladed turbine.

\begin{figure}
\begin{subfigure}[t]{0.48\textwidth}
\centering
  \includegraphics[width=\textwidth]{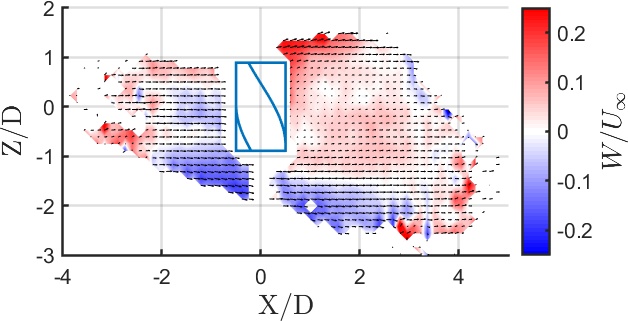}
  \caption{}
\label{fig:w_UGE}
\end{subfigure}
\hfill
\begin{subfigure}[t]{0.48\textwidth}
\centering
  \includegraphics[width=\textwidth]{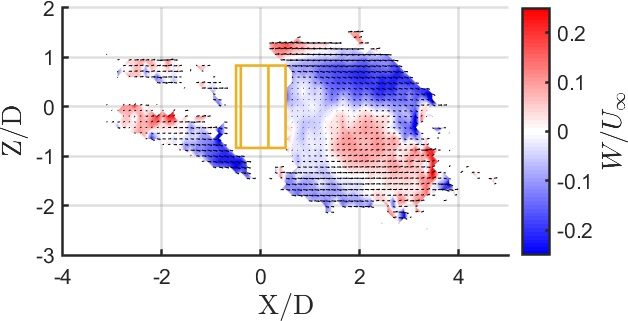}
  \caption{}
\label{fig:w_WPE}
\end{subfigure}
\caption{Time-averaged planar fields of the vertical velocity $W$ for (a) the helical-bladed turbine at $\lambda = 1.19$ and (b) the straight-bladed turbine at $\lambda = 1.20$, taken at $Y/D=0$. The wake of the straight-bladed turbine is characterized by symmetric sweeps of high-momentum fluid into the wake from above and below. In contrast, the wake of the helical-bladed turbine exhibits a uniform updraft at $Y/D=0$. This difference suggests that the helical blades have a pronounced three-dimensional effect on the wake structure.}
\label{fig:w}
\end{figure}

Secondly, a slice of the time-averaged vertical-velocity component $W$ through the central axis of the turbine further implied the existence of more complex three-dimensional dynamics in the wake of the helical-bladed turbine (figure \ref{fig:w}). The vector field at $Y/D=0$ for the straight-bladed turbine showed a downward sweep of fluid from above the turbine and an upward sweep of fluid from below the turbine, as would be expected from a bluff body in crossflow (figure \ref{fig:w_WPE}). The corresponding field for the helical-bladed turbine showed a very different scenario, in which a uniform central updraft was present (figure \ref{fig:w_UGE}). At planar slices of $Y/D$ on either side of $Y/D=0$, corresponding uniform downward motions of fluid were observed. These differences in the vertical velocity fields implied that the 3D structure of the helical-bladed turbine had an effect on the wake dynamics that could not be resolved simply by examining the velocity fields. A three-dimensional analysis of the vortical structures present in the wakes of these VAWTs is required to explain these observed differences. This will be provided in section \ref{sec:sec3_vorticities}.

\subsection{Vortical Structures and Wake Topology}
\label{sec:sec3_vorticities}

\begin{figure}
\centering
\includegraphics[width=\textwidth]{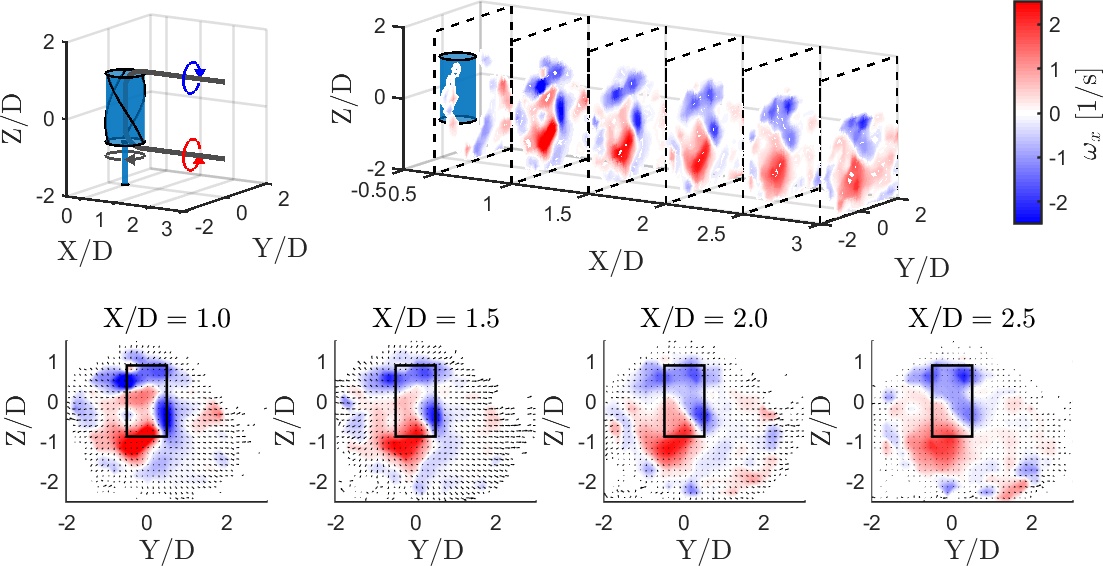}
\caption{Streamwise slices of the streamwise vorticity $\omega_x$ in the case of the helical-bladed turbine for $\lambda = 1.19$. The $X$-axis is stretched on $0.5\leq X/D \leq 3$ to show the slices more clearly. These fields show marked asymmetry and a vertical misalignment in the two branches of the horseshoe vortex induced by the rotation of the turbine, compared to those shown in figure \ref{fig:wx_WPE}.}
\label{fig:wx_UGE}
\end{figure}

\begin{figure}
\centering
\includegraphics[width=\textwidth]{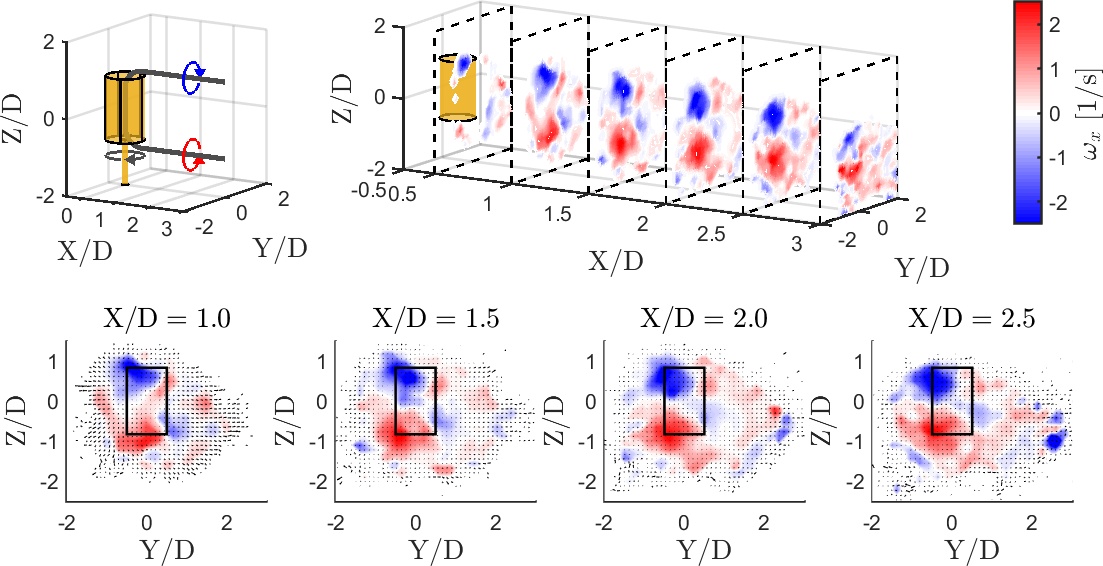}
\caption{Streamwise slices of the streamwise vorticity $\omega_x$ in the case of the straight-bladed turbine for $\lambda = 1.20$. The $X$-axis is stretched on $0.5\leq X/D \leq 3$ to show the slices more clearly. Compared to the wake of the helical-bladed turbine (figure \ref{fig:wx_UGE}), the streamwise vortical structures are symmetric about the $Z/D=0$ plane. Small counter-rotating secondary vortices are also present to the right of each main streamwise vortex, {possibly similar to those observed in full-scale HAWTs by \cite{yang_coherent_2016}}.}
\label{fig:wx_WPE}
\end{figure}

An analysis of the vortical structures in the streamwise direction ($\omega_x$) demonstrates the importance of three-dimensional considerations to the wake dynamics of these VAWTs. Streamwise planar slices of $\omega_x$ are shown in figures \ref{fig:wx_UGE} and \ref{fig:wx_WPE}. The structures visible in these plots comprised a horseshoe-shaped vortex induced by the rotation of the turbine, observed previously by \cite{brownstein_aerodynamically_2019}. While in the case of the straight-bladed turbine, the two branches of the vortex were symmetric about the mid-span of the turbine, the corresponding structure for the helical-bladed turbine was asymmetric. The offset of the upper branch with respect to the lower branch induced the central updraft of fluid observed in figure \ref{fig:w_UGE}. It will be argued in section \ref{sec:sec3_twist} that this asymmetry was a result of the blade twist of the helical-bladed turbine, which skewed the overall wake profile and thereby affected the alignment of the streamwise vortical structures.

\begin{figure}
\centering
\includegraphics[width=\textwidth]{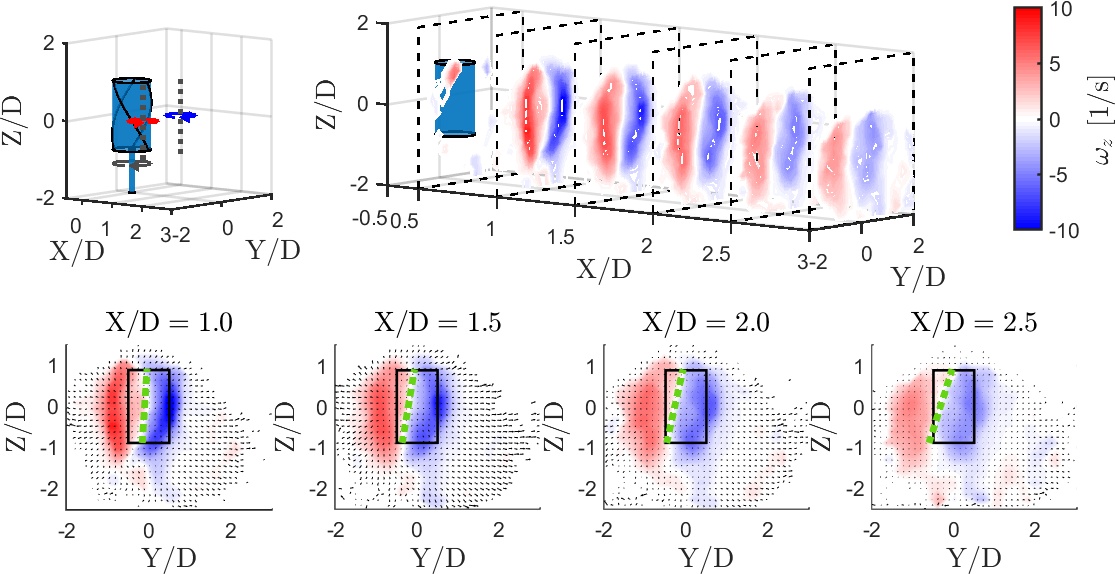}
\caption{Streamwise slices of the vertical vorticity $\omega_z$ downstream of the helical-bladed turbine for $\lambda = 1.19$. As in the previous figures, the $X$-axis is stretched on $0.5\leq X/D \leq 3$. These structures exhibit a tendency to tilt with increasing streamwise distance from the turbine, {as evidenced by fits to the zero-vorticity region between the structures (dashed green lines)}.}
\label{fig:wz_UGE}
\end{figure}

\begin{figure}
\centering
\includegraphics[width=\textwidth]{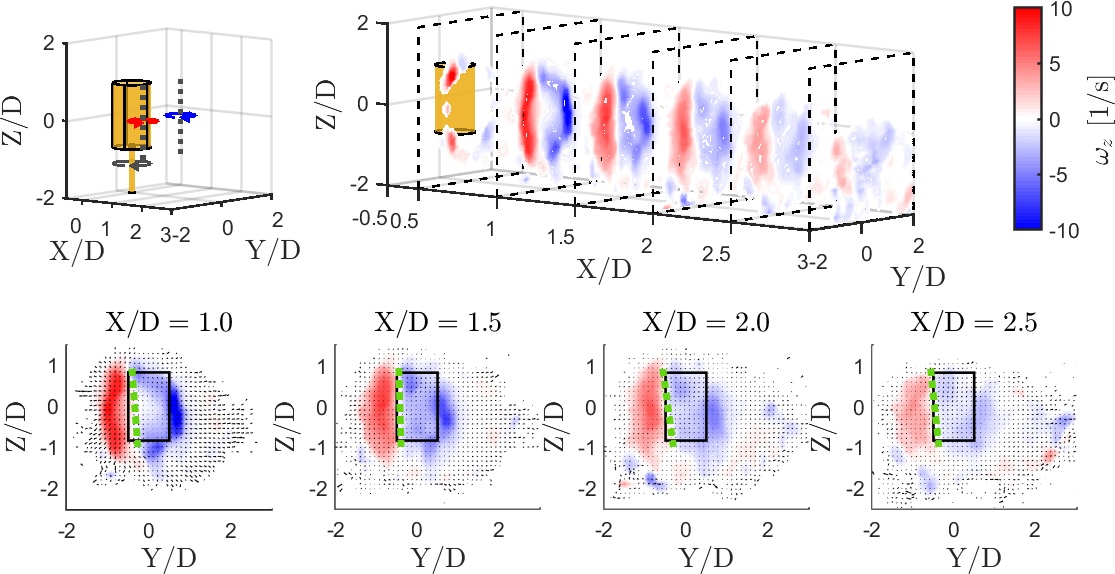}
\caption{Streamwise slices of the vertical vorticity $\omega_z$ downstream of the straight-bladed turbine for $\lambda = 1.20$. The $X$-axis is again stretched on $0.5\leq X/D \leq 3$. These structures remain upright with respect to the vertical {(again denoted by dashed green lines)}, in contrast to their counterparts from the helical-bladed turbine.}
\label{fig:wz_WPE}
\end{figure}

Similar asymmetric behavior in the wake of the helical-bladed turbine was observed in the vertical vortical structures ($\omega_z$), shown in streamwise slices in figures \ref{fig:wz_UGE} and \ref{fig:wz_WPE}. In both cases, the structure with positively signed vorticity initially had a linear shape and was oriented vertically, while the structure with negatively signed vorticity was bent toward the positive $Y$ direction. This initial geometry was related to the mechanics of formation of these vortices. The positively signed structure was composed of vortices shed from turbine blades as they rotated into the wind, which formed a vortex line that was advected downstream. These vortices have been observed in 2D for straight-bladed turbines by \cite{tescione_near_2014}, \cite{parker_effect_2016}, and \cite{araya_transition_2017}, {and in 3D simulations of straight-bladed turbines by \cite{villeneuve_improving_2020}}. The arched shape of the negatively signed structures was a consequence of the low-velocity wake region. Behind the straight-bladed turbine, these two structures remained vertically oriented and relatively parallel. In contrast, behind the helical-bladed turbine, these structures began to tilt with respect to the vertical as they were advected downstream. This tilt was analogous to that observed in the horseshoe vortex (figure \ref{fig:wx_UGE}). Since the structures in $\omega_z$ were formed by vortex shedding from the turbine blades, we hypothesize that the helical blades of the UGE turbine were the cause of this observed asymmetric wake behavior.

The spanwise vortical structures ($\omega_y$) did not show any significant signs of asymmetry (appendix \ref{sec:appC_vorticity}, figures \ref{fig:wy_UGE} and \ref{fig:wy_WPE}). These structures represented time-averaged tip vortices shed by the passing turbine blades, as documented by \cite{tescione_near_2014}, and were thus not expected to change in geometry in these experiments. Together, the blade-shedding structures in $\omega_y$ and $\omega_z$ bounded the near wake. Their time-averaged profiles outlined and encapsulated the regions of streamwise-velocity deficit in the wake shown in figures \ref{fig:u_UGE} and \ref{fig:u_WPE}. This topological correspondence suggested that vortex shedding from the turbine blades has a dominant effect on the overall shape of the near wake.


\subsection{Effect of Blade Twist}
\label{sec:sec3_twist}

In the previous section, connections between vortex shedding from VAWT blades and the 3D topology of VAWT wakes were observed. A more thorough investigation of the tilted wake is now undertaken to develop a more comprehensive description of the dynamics in the near wake.

\begin{figure}
\begin{subfigure}[t]{0.48\textwidth}
\centering
  \includegraphics[width=\textwidth]{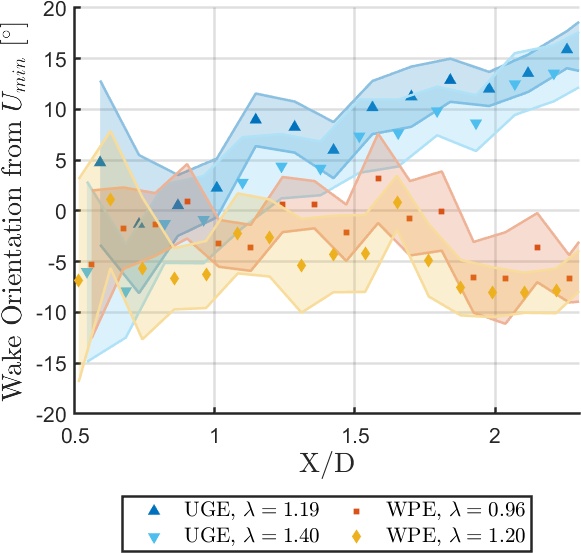}
  \caption{}
\label{fig:orientation_vel}
\end{subfigure}
\hfill
\begin{subfigure}[t]{0.48\textwidth}
\centering
  \includegraphics[width=\textwidth]{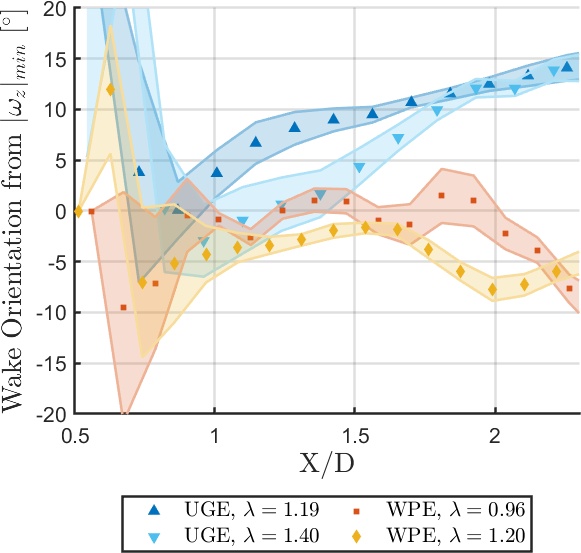}
  \caption{}
\label{fig:orientation_wz}
\end{subfigure}
\caption{Wake orientation measurements from all four experimental cases, computed from (a) the locations of minima in the velocity-deficit region, $U_{min}$, and (b) the coordinates of the zero-vorticity strip between the two vertical vortical structures, $|\omega_z|_{min}$. The wake orientation of the helical-bladed turbine increases monotonically, while that of the straight-bladed turbine does not exhibit a strong trend away from zero.}
\label{fig:orientation}
\end{figure}

The results presented thus far have shown that the wake of the helical-bladed turbine was tilted at some angle with respect to the vertical, whereas the wake of the straight-bladed turbine was not. To quantify this effect, two measures were employed: the angle of the velocity-deficit region, and the angle of the region of zero vorticity between the two vertical vortical structures. These two measures were selected on the premise that the dynamics of the vertical vortical structures are tied to the geometry of the near wake. Both were computed on slices parallel to the $YZ$ plane, taken at several streamwise positions downstream of each turbine. For the first measure, the location of minimum velocity was identified at every $Z$-position in each slice, and a linear fit through these points on each slice was computed to approximate the slope of the velocity-deficit region. For the second measure, the location of minimum vorticity between the two vertical vortical structures was identified at every $Z$-position in each slice using linear interpolation, and a linear fit through these points on each slice represented the orientation of the structures. For both measures, confidence intervals of one standard deviation on the slope of the linear fit served as error bounds. The results of this procedure are shown in figure \ref{fig:orientation_vel} for the velocity-deficit measure and figure \ref{fig:orientation_wz} for the vortical-structures measure. The results demonstrated that the wake orientation of the straight-bladed turbine did not exhibit a strong deviation from the vertical in the near-wake region. In contrast, the wake orientation of the helical-bladed turbine increased monotonically with streamwise distance for both of the tip-speed ratios tested in the experiments. These measures thus quantified the various observations from the previous section regarding changes in the wake topology between the two turbines.

The wake-orientation measurements shed light on the mechanism responsible for the tilted wake observed for the helical-bladed turbine. First, the orientation profiles were very consistent between the two measures, supporting the hypothesis that the shape of the near wake is directly tied to the dynamics of the vertical vortical structures. Additionally, the profiles did not show a strong dependence on turbine solidity, {and no significant dependence on tip-speed ratio over the limited range of $\lambda$ tested in these experiments was observed}. Though the measurements did separate into two classes corresponding to the two turbine geometries, the three-dimensional nature of the tilted-wake behavior made it unlikely that solidity, which is not defined using three-dimensional geometric parameters, was the dominant factor. These results thus isolate the geometry of the turbine blades as the primary contributing factor to the observed differences in wake topology.

A mechanism by which turbine-blade geometry can affect the wake topology is now proposed. As described previously, the vortical structures visible in the time-averaged fields of $\omega_z$ represent the profiles of vortex lines shed from the rotating turbine blades and advected away from the turbine by the free stream. We expect that the shape of these vortex lines will depend on the shape of the blades: a straight blade will shed a straight vortex line, while a helical blade will shed a vortex line with nonzero curvature. The latter hypothesis is drawn from the fact that a helical blade, in contrast to a straight blade, is yawed relative to the incoming flow and experiences a range of local angles of attack along its span as it rotates around the turbine. A yawed or swept blade with respect to the incoming flow exhibits strongly three-dimensional vortex shedding in dynamic stall \citep{visbal_effect_2019}, due in part to a net transport of vorticity along the span of the blade \citep{smith_measurements_2019}. A linearly increasing spanwise angle-of-attack profile in dynamic stall similarly induces a spanwise transport of vorticity that affects the stability of the leading-edge vortex and thus the character of the vortex shedding from the blade \citep{wong_flow_2017}. Given these observations, we conclude that the spanwise nonuniformity of the flow over helical VAWT blades makes the vortex lines shed by the dynamic-stall mechanism inherently nonlinear and three-dimensional.

To model the evolution of these vortex lines as they are advected downstream, the principle of Biot-Savart self-induction can be applied. In an inviscid flow field, the self-induced velocity at any point $\mathbf{r}$ on a single vortex line with finite core size $\mu$, defined by the curve $\mathbf{r'}$ and parameterized by the arc length $s'$, can be written as 

\begin{equation}
\frac{\partial\mathbf{r}}{\partial t} = -\frac{\Gamma}{4\pi} \int \frac{(\mathbf{r}-\mathbf{r'})\times \frac{\partial \mathbf{r'}}{\partial s'}}{(\left|\mathbf{r}-\mathbf{r'}\right|^2+\mu^2)^{3/2}} ds',
    \label{eqn:biotsavart}
\end{equation}

\noindent where the integration is performed over the length of the vortex line \citep{leonard_computing_1985}. This model for the self-induced deformation of curved vortex lines has been studied numerically using both approximate methods \citep{arms_localizedinduction_1965} and exact simulations \citep{moin_evolution_1986}. The quantity $(\mathbf{r}-\mathbf{r'}) \times \frac{\partial \mathbf{r'}}{\partial s'}$ in the numerator of the integrand is only nonzero when the displacement vector between two points on the vortex line does not align with the direction of the vortex line. Therefore, a vortex line with curvature or piecewise changes in alignment will undergo deformation under self-induction, while a purely linear vortex line will not. Self-induced deformations of a similar nature have been observed in curved and tilted vortex lines in numerous computational and experimental contexts \cite[e.g.][]{hama_selfinduced_1961,boulanger_tilt-induced_2008}.

\begin{figure}
\centering
\includegraphics[width=0.48\textwidth]{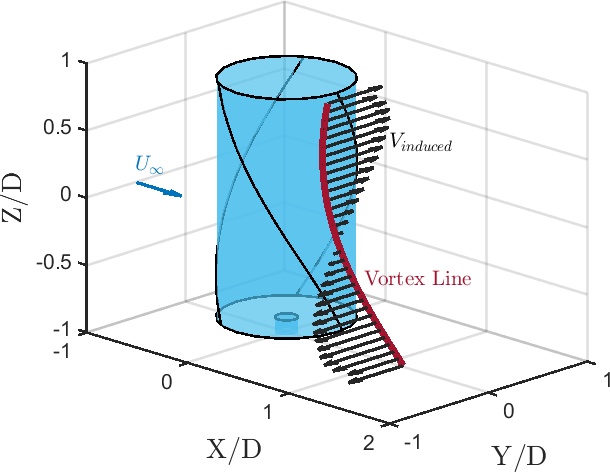}
\caption{Schematic of the $Y$ component of the induced velocities, $V_{induced}$, along a helical vortex line due to Biot-Savart self-induction (equation \ref{eqn:biotsavart}). The scale of the vectors and the streamwise location of the vortex line are both arbitrary, and the streamwise and vertical components of the induced velocity are not shown for clarity. The stretching induced on the vortex line matches the behavior of the tilted wake.}
\label{fig:BS}
\end{figure}

Given that the precise shape of the curved vortex lines shed by the helical-bladed turbine cannot be extracted from the time-averaged vorticity fields, equation \ref{eqn:biotsavart} cannot be applied quantitatively in this case. However, it can still be used qualitatively to connect the development of the tilted wake to blade geometry. We therefore consider a helical vortex line that corresponds to the helical shape of the UGE turbine blades, shed from a blade as it rotates upstream into the prevailing wind. This model system accounts for the three-dimensional blade geometry while abstracting the precise dynamics of the vortex-shedding mechanism on the blades. The curve is parameterized for $Z \in \left[-S/2, S/2\right]$ as $X = -a R \sin(\tau Z)$ and $Y = - R \cos(\tau Z)$. The constant $a$ accounts for stretching of the vortex line in the streamwise direction due to differences between the tip-speed velocity of the turbine and advection from the free stream. The selection $a = 1/\lambda$, for example, recovers the expected asymptotic result that the curved vortex line will become a straight vertical line as $\lambda \rightarrow \infty$. The initial induced velocities along this vortex line, computed numerically from equation \ref{eqn:biotsavart}, apply a stretching in the $Y$ direction that corresponds directly with the previous observations of the tilted wake (figure \ref{fig:BS}). The streamwise and vertical induced velocities are not addressed in this analysis since they do not contribute to the tilted wake. The computed vectors also do not represent the full time-evolution of the vortex line, as only the initial induced velocities $\mathbf{u}\left(\mathbf{r}(t=0)\right)$ are given. The demonstration shows qualitatively that the mechanism of Biot-Savart self-induction provides a direct connection between blade geometry and the evolution of the wake topology.

Based on these results, the propagation of the influence of turbine-blade geometry in the dynamics of the near wake can be outlined. The shape of the blades determines the shape of the vortex lines shed by the blades as they rotate around the turbine. The shape of these vortex lines in turn drives their evolution downstream of the turbine. As these structures bound the wake region, topological changes in the vortex lines are reflected in the shape of the velocity-deficit region of the wake. These developments affect the shape of the horseshoe vortex, which is not directly affected by differences in blade geometry but is necessarily bound to the shape of the wake region. The connection between blade twist and wake tilt described in this section therefore provides a unifying framework that accounts for the trends previously observed in the wake velocity and vorticity fields.

\subsection{Wake Dynamics}
\label{sec:sec3_dynamics}

\begin{figure}
\centering
\includegraphics[width=0.5\textwidth]{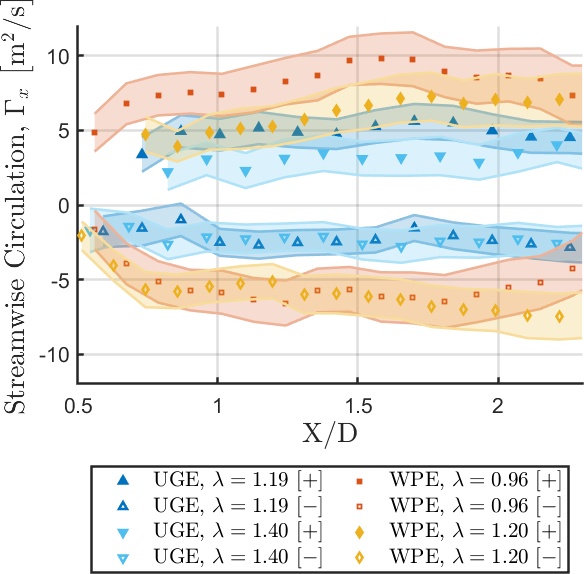}
\caption{Circulation of the positive and negative streamwise vortical structures, $\Gamma_x$, for all four experimental cases. Compared to the plots of $\Gamma_y$ and $\Gamma_z$ shown in figure \ref{fig:circulation_yz}, $\Gamma_x$ did not decay as significantly with increasing streamwise distance, and the horseshoe vortex was thus hypothesized to extend farther into the wake than the structures induced by vortex shedding from the blades.}
\label{fig:circulation_x}
\end{figure}

\begin{figure}
\begin{subfigure}[t]{0.48\textwidth}
\centering
  \includegraphics[width=\textwidth]{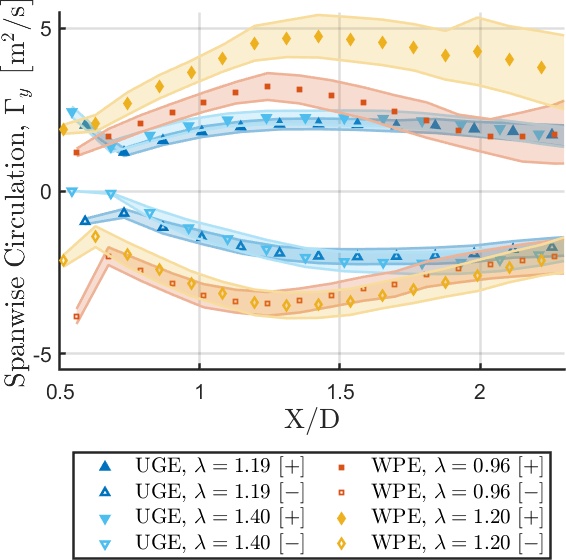}
  \caption{}
\label{fig:circulation_y}
\end{subfigure}
\hfill
\begin{subfigure}[t]{0.48\textwidth}
\centering
  \includegraphics[width=\textwidth]{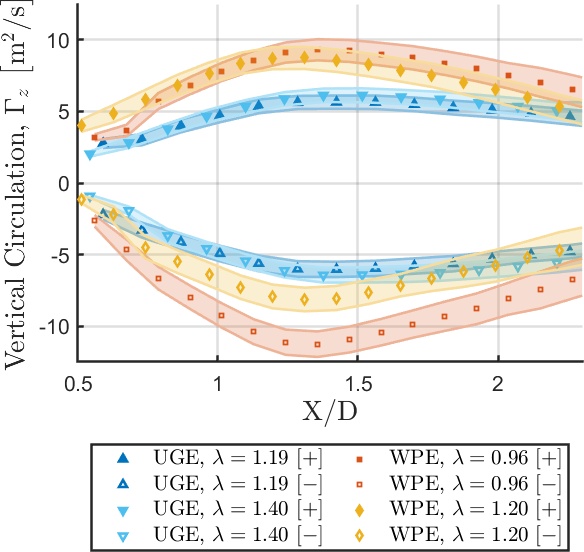}
  \caption{}
\label{fig:circulation_z}
\end{subfigure}
\caption{Circulations of the positive and negative (a) spanwise vortical structures ($\Gamma_y$) and (b) vertical vortical structures ($\Gamma_z$) for all four experimental cases. The circulation profiles collapse approximately by turbine, implying that turbine solidity is a dominant factor in these dynamics. The decaying trends of the profiles past $X/D \approx 1.5$ show that these structures are only influential in the near wake.}
\label{fig:circulation_yz}
\end{figure}

In the previous section, it was hypothesized that the shape of VAWT blades dictates the near-wake topology through Biot-Savart self-induction of shed vortex lines. The wake dynamics are now analyzed in detail, to quantify the evolution of the vortical structures in the wake and to identify their contributions to wake recovery. The circulation of the wake structures in each direction, $\Gamma_i$, was calculated as a function of downstream distance from the turbine. The circulation was computed at a series of streamwise positions $X/D$ by integrating the vorticity component in question over a square $D \times D$ window, which was oriented normal to the direction of the vorticity component. This window was placed at the center of the vortex at several points along the vortex line. The center was identified at each point by applying a threshold based on the standard deviation of the vorticity field and computing the center of mass of the isolated vorticity distribution. The values of the circulation along the vortex line, computed with these windows, were averaged to obtain a representative circulation for the structure. Error bars were determined by propagating the standard deviations of the components of individual velocity vectors from the particle trajectories in each voxel through the curl operator and the circulation integration. This computation was done independently for the positively and negatively signed vortical structures. The resulting circulations in $X$, $Y$, and $Z$ are plotted on $0.5 \leq X/D \leq 2.3$ in figures \ref{fig:circulation_x}, \ref{fig:circulation_y}, and \ref{fig:circulation_z}, respectively. {Circulation measurements downstream of this region were inconclusive due to measurement noise, or possibly as a result of atmosphere-induced unsteady modulations in the wake similar to those observed in full-scale HAWTs \citep{abraham_dynamic_2020}.}


The circulations of the streamwise structures ($\Gamma_x$, {figure \ref{fig:circulation_x}}) appeared to extend farther into the wake than the vortical structures in $Y$ and $Z$, whose circulations declined monotonically after $X/D\approx1.5$ ({as shown in} figure \ref{fig:circulation_yz}). This difference in dynamical behavior agrees with the previously stated hypothesis that the horseshoe vortex is induced by the rotation of the turbine rather than unsteady vortex shedding from the turbine blades. Furthermore, the streamwise circulation at the lowest tip-speed ratio ($\lambda = 0.96$) began to decline at $X/D \approx 1.7$, prior to those at higher tip-speed ratios, which suggests that the persistence of the horseshoe vortex in the wake increases with increasing $\lambda$. Given the limited range of tip-speed ratios and streamwise locations considered in these experiments, a definitive relation between $\Gamma_x$ and $\lambda$ cannot be isolated from these results. Still, the comparatively long-lived coherence of the horseshoe vortex in the wake does imply that it plays an important role in the dynamics of wake recovery, considering its influence on vertical velocities in the wake and thus the entrainment of momentum into the wake (figure \ref{fig:w}). Additionally, the difference in the alignment of the branches of the horseshoe vortex between the helical- and straight-bladed turbines suggests that the wake recovery will also differ as a function of blade geometry.

The circulations of the structures corresponding to vortex shedding from the turbine blades ($\Gamma_y$ and $\Gamma_z$, shown in figures \ref{fig:circulation_y} and \ref{fig:circulation_z}) exhibited uniform behavior for both turbines. Taken together with the time-resolved results of \cite{araya_transition_2017}, these trends confirm that the vortex-shedding dynamics that contribute to the wake structure are only active in shaping the near wake ($X/D\lesssim2$). An apparent dependence on turbine solidity is evident in these data, as the circulation profiles separated into groups by turbine rather than by tip-speed ratio. This trend is likely not a function of three-dimensional blade shape, because the measurements of $\Gamma_y$ and $\Gamma_z$ were averaged in $Y$ and $Z$, respectively. Again, because of the limited number of turbine geometries considered in this study, a definitive relation between $\Gamma_y$, $\Gamma_z$, and $\sigma$ cannot be established from these data. These results, however, do indicate that the contributions of the blade-shedding structures to wake recovery will not differ significantly between the two turbine geometries considered in this study.

The main conclusion from the circulation analysis presented in this section is that the horseshoe vortex extends farther into the wake than the vortical structures in the $Y$ and $Z$ directions, which begin to decay in strength soon after they are shed from the turbine blades. These results thus suggest that, though the vortex lines shed by the turbine blades define the topology of the near wake, the horseshoe vortex has a larger contribution to wake recovery due to its comparative longevity in the wake. The analysis also noted {possible} dependencies of $\Gamma_x$ on $\lambda$ and of $\Gamma_y$ and $\Gamma_z$ on $\sigma$. The contributions of vortex breakdown and turbulent entrainment of momentum from the free stream to wake recovery have not been considered here, because of the time-averaged nature of the experimental data. These effects become more significant downstream of the limit of our measurement domain, and thus will also be affected by the mean-flow dynamics identified in this work.

\subsection{Implications for Turbine Arrays}
\label{sec:sec3_implications}

The findings of this study have parallels in recent studies of the wake dynamics of HAWTs that highlight the implications of the present results for the arrangement of VAWTs in closely packed arrays. In both experiments with a porous disc and large-eddy simulations with actuator-disc and actuator-line models, \cite{howland_wake_2016} reported an alteration to the shape of the wake profile, known as a curled wake, behind HAWTs yawed with respect to the incoming flow caused by counter-rotating streamwise vortices that are induced by the yawed turbine. \cite{bastankhah_experimental_2016} and \cite{shapiro_modelling_2018} modelled this topological phenomenon analytically, leading to new estimates of the wake deflection and available power downwind of a yaw misalignment wind turbine. \cite{fleming_full-scale_2017} confirmed the existence of these vortices for yawed turbines in field conditions. Additionally, \cite{fleming_simulation_2018} demonstrated that these changes in wake topology affect the performance of downstream and adjacent turbines in an array. \cite{howland_wind_2019} and \cite{fleming_initial_2019} then utilized these analytic models to demonstrate the potential for wake steering control in field experiments of HAWTs at full scale. Given the qualitative similarities in the dynamics of the tilted wake observed in this study and the curled wake observed in HAWTs, the ramifications of the tilted wake for arrays of VAWTs could be analogous as well.

The performance of VAWTs downstream of a given turbine will be affected by the dynamics identified in this study. The strong vortex shedding in the near wake of a VAWT ($X/D \lesssim 2$) will lead to a severe decrease in the performance of downstream turbines placed in this region. \cite{brownstein_aerodynamically_2019} observed this effect in experiments with turbine pairs. For greater streamwise separations, the horseshoe vortex will be the primary large-scale mean-flow vortical structure encountered by downstream turbines. At these distances, the effects of helical blades on wake recovery will be more evident, and will affect the overall performance of the VAWT array. The tilted wake will also lead to a modified wake profile that will affect the optimal placement of helical-bladed turbines in arrays {by at least one turbine radius}.

The performance of adjacent VAWTs will also be affected, especially for turbines placed in close proximity to each other for the enhancement of power production. The mean-flow mechanisms observed by \cite{brownstein_aerodynamically_2019} will be altered by the presence of the three-dimensional vortex shedding from helical-bladed turbines, and it is thus likely that these turbines would experience a different degree of performance enhancement observed in that work for straight-bladed turbines.

Lastly, the vortical structures and topological effects isolated in this study are expected to hold for vertical-axis wind turbines in general, with variations for different turbine geometries. The horseshoe vortex, in particular, is expected to be enhanced for VAWTs that operate at higher tip-speed ratios than those employed in these experiments. The significance of the tilted wake for helical-bladed turbines will vary with aspect ratio and blade twist. The 3D effects of blade geometry will also need to be reconsidered for the bowed blades of the large-scale Darrieus-type turbines studied by \cite{klimas_effects_1981}. Particular differences notwithstanding, it can be inferred that the three-dimensional dynamics identified here will have significant effects on the dynamics and topology of the wake for VAWTs of all designs and power outputs, and thereby affect the aerodynamics and overall efficiency of VAWT arrays.

\section{Conclusions}

In this study, {a novel method for conducting volumetric three-dimensional, three-component flow-field measurements in field conditions was developed and used to quantify the wake dynamics of} two full-scale vertical-axis wind turbines. {This 3D-PTV setup demonstrated sufficient precision and resolution to resolve large-scale vortical structures in the wakes of the turbines. Time-averaged} velocity and vorticity fields showed a tilted wake downstream of helical-bladed turbines, compared to the wake profile of straight-bladed turbines. A fully three-dimensional analysis of the vortical structures present in the near wakes of VAWTs was undertaken, and demonstrated that the topology of the near wake is dependent primarily on the mechanics of vortex shedding from the turbine blades. The connection between blade geometry and wake dynamics was clarified by considering the Biot-Savart self-induction of curved vortex lines. Measurements of the circulation of the vortical structures in the wake revealed that the vortical structures shed by the blades decay in influence relatively quickly but affect the topology of the horseshoe vortex, which extends farther into the wake and is thus related to wake recovery. Therefore, a line of influence was established between blade geometry, vortex shedding, near-wake topology, the horseshoe vortex, and wake recovery.

A major point of emphasis in this work is that three-dimensional effects are significant to the dynamics and evolution of VAWT wakes. Many of the flow phenomena observed in this study would be difficult to resolve completely by planar measurement techniques; the vorticity and circulation measurements in particular required three-component 3D data to collect. The experimental method for obtaining time-averaged velocity and vorticity measurements in field conditions and high Reynolds numbers is therefore well suited to further analyses of wind-turbine wake dynamics at full scale. {The method is also flexible enough in its deployment to be applied to other large-scale flow phenomena and atmosphere-structure interactions.}

While this study focused on differences in flow topology that stemmed from the use of helical blades in VAWTs, the connection between object geometry and vortex dynamics proposed in this work could be applied in wider aerodynamic contexts, such as rotorcraft, flapping flight, and more complex fluid-structure interaction problems. This study provides both an experimental method and a theoretical framework for wake analysis that can be leveraged for future studies of the three-dimensional vortex dynamics in engineering-scale wakes.

The limited number of test cases in this study precluded the execution of a full scaling analysis of the vortical structures and circulation profiles with respect to turbine solidity and tip-speed ratio. A larger series of test cases, involving different turbine geometries and a wider range of tip-speed ratios, would allow these relationships to be established more quantitatively. In addition, a larger measurement volume would allow the wake-topology observations of this study to be extended into the far wake, so that the effects of the tilted wake and the streamwise vortical structures on wake recovery could be more thoroughly investigated. Given the size of the measurement domain that would be required for these kinds of experiments, as well as the range of parameters involved, it may be more feasible to carry out this analysis in a laboratory or computational setting. This study can thus provide validation cases to demonstrate that flow phenomena observed in VAWT wakes in laboratory experiments and numerical simulations are representative of those present in the wake of operational turbines in field conditions.

\section*{Acknowledgements}

The authors gratefully acknowledge funding from the Gordon and Betty Moore Foundation
through Grant No. 2645, the National Science Foundation through Grant No. FD-1802476, the Stanford University TomKat Center for Energy Sustainability, and the Stanford Graduate Fellowships in Science and Engineering. The authors would also like to recognize Bob Hayes and Prevailing Wind Power for managing the operation and maintenance of the FLOWE field site, and Ryan McMullen for assisting with the field experiments. {Lastly, the authors extend their thanks to Snow Business International \& Snow Business Hollywood for assisting in the selection and supply of the snow machines and the snow fluid used, for providing troubleshooting during late night experiments, for allowing extended use of their machines, and for shipping a machine to Stanford for the laboratory experiments.}

\section*{Declaration of Interests}

The authors report no conflict of interest.

\appendix

\section{Dynamics of artificial snow Particles}
\label{sec:appendix_snow}

Since the choice of {flow seeding} particles is critical to the accuracy of PTV measurements, the generation and characterization of the artificial snow particles used in these experiments were carefully considered. Natural snowfall had been used successfully by \cite{hong_natural_2014} as {seeding} particles for 2D particle-image velocimetry in the wake of a full-scale HAWT. In the present experiments, artificial snow was used due to the lack of natural snowfall at the FLOWE site. The artificial snow used in these experiments was composed of an air-filled soap foam {(ProFlake Falling Snow Fluid, Snow Business)}, which formed irregularly shaped particles {with a range of sizes. The particles that were visible in the images from the field site had diameters of $d_p = 11.2\pm4.2$ mm and an average density of $\rho=6.57\pm0.32$ $\rm{kgm^{-3}}$}. Smaller particles would have been more ideal as {`tracer' particles}, but these were difficult to visually identify and isolate in the recorded videos and were thus unfeasible for these experiments. The settling velocity of these particles in quiescent air was measured to be $W_s = 0.60 \pm 0.18$ $\rm{ms^{-1}}$, which corresponds to a particle Reynolds number $Re_p = \frac{W_s d_p}{\nu}$ of {448}. Since this Reynolds number is outside the Stokes-flow regime, the relative influence of inertial effects on the ability of the particles to follow the flow had to be ascertained. {Empirical correlations for general airborne particles \cite[e.g.][]{bohm_general_1988} failed to predict the dynamics of the artificial snow particles because of their foam-based composition, and therefore laboratory experiments were required.}

{In this section, a series of laboratory experiments are detailed that allowed the flow response of the artificial snow particles to be quantified. This is followed by a qualitative analysis based on scaling arguments that confirms the findings of the laboratory experiments in the field data.}

\subsection{Laboratory Experiments with a Snow Machine}
\label{sec:appA_experiments}

{Laboratory experiments in a wind tunnel were designed to determine the aerodynamic response characteristics of the artificial snow particles used in the field experiments. To measure the particle response to a step change in velocity, a snow machine was arranged to release particles normal to the flow in the wind tunnel, creating a jet in crossflow. The spanwise discrepancy between the profile of the jet in crossflow and the particle trajectories could then be analyzed to establish a particle relaxation time and quantify the effects of particle inertia on flow fidelity. The dynamics of the large particles used in the field experiments were therefore compared with the behavior of smaller, more regularly shaped particles generated by increasing the blower flow rate and decreasing the fluid injection rate in the snow machine. The response of each particle type was studied at two distinct free-stream velocities. The impulse response and slip velocity of the particles were then calculated from the measured jet profiles.}

{Experiments to characterise the dynamics of particles produced by a single snow machine (Silent Storm DMX, Ultratec Special Effects) were conducted in an open-circuit wind tunnel with flow driven by a $4\times4$ grid of fans at the inlet of the tunnel. The test section was 4.88 m long in the streamwise direction ($X$), 2.06 m in width ($Y$), and 1.97 m in height ($Z$). The tunnel was operated with free-stream velocities of $5.64\pm0.45$ $\rm{ms^{-1}}$ and  $6.58\pm0.45$ $\rm{ms^{-1}}$, as measured with a hot-film anemometer. A full description of the wind-tunnel facility is given by \cite{brownstein_aerodynamically_2019}.}

\begin{figure}
\centering
\includegraphics[width=0.5\textwidth]{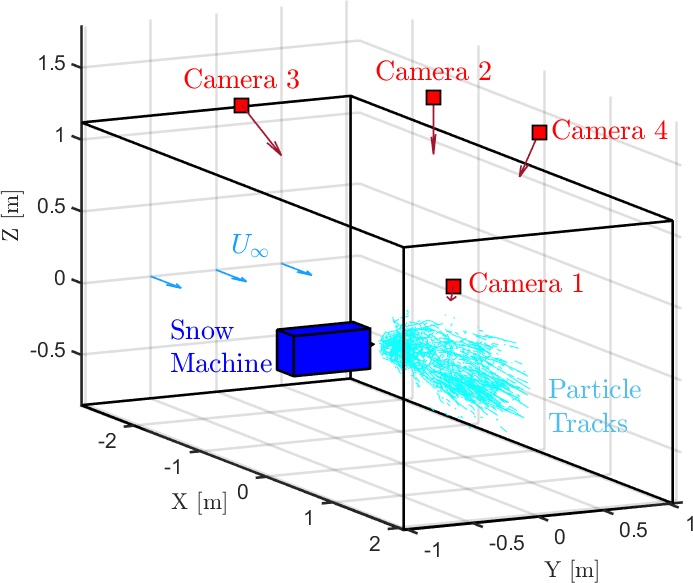}
\caption{{Schematic of the experimental setup for snow-machine experiments. Flow is in the positive $X$ direction, and the snow machine (blue) emits particles in the positive $Y$ direction. Sample particle tracks (light blue) illustrate the extent of the measurement volume. The four cameras are shown in red, and arrows denote their viewing angles.}}
\label{fig:expSetup_snow}
\end{figure}

{The snow machine was positioned on a platform 2.8 m downstream of the tunnel inlet, so that the output nozzle was perpendicular to the centerline of the tunnel (cf.\ figure \ref{fig:expSetup_snow}). The end of the nozzle was offset at $Y = -0.2$ m in the spanwise direction from the center of the tunnel. The blockage from this configuration was less than 5\%. Four hardware-synchronized cameras (N-5A100, Adimec) were positioned around the tunnel, with one located directly above the measurement volume. These cameras captured a measurement volume of approximately 1.2 m $\times$ 0.5 m $\times$ 0.5 m in the $X$, $Y$, and $Z$ directions, respectively. The cameras recorded images at a resolution of $2048\times1008$ pixels and a frequency of 250 frames per second, with an exposure of 600 $\rm{\mu s}$.}

\begin{figure}
\centering
\includegraphics[width=1\textwidth]{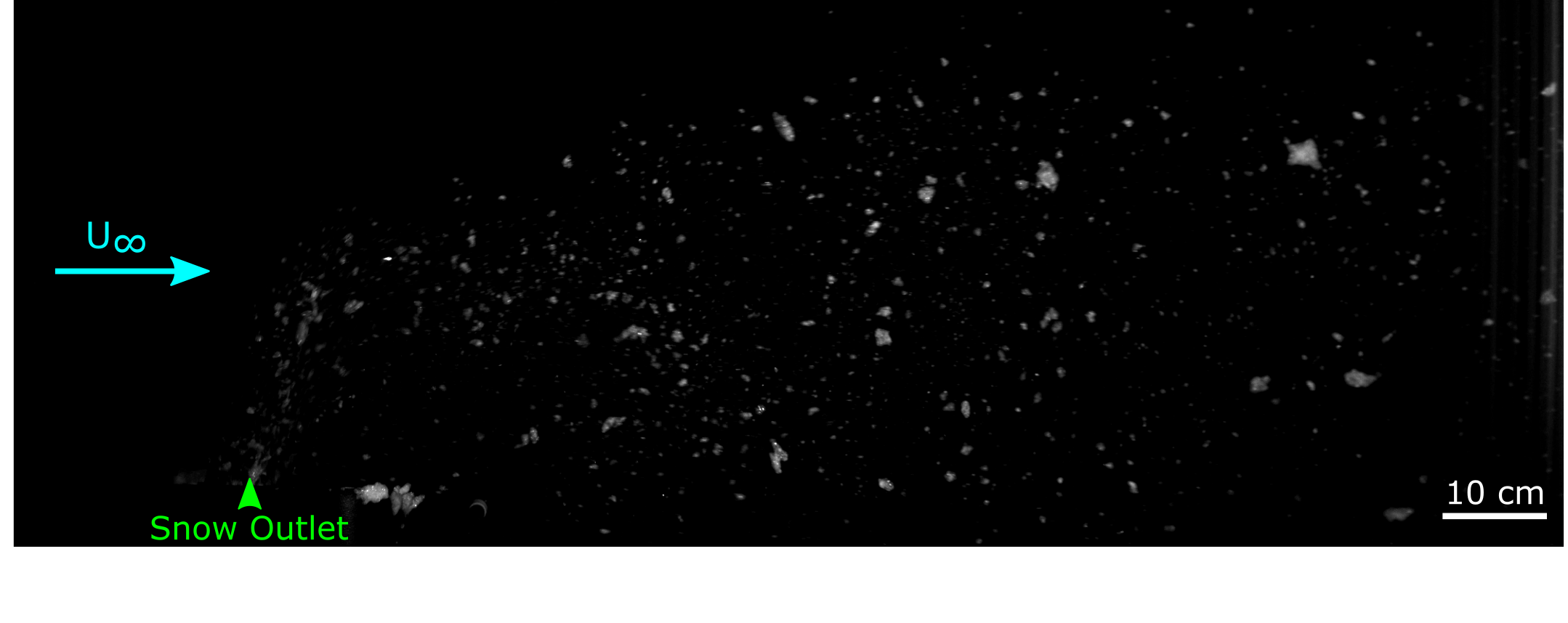}
\caption{{Photograph of snow particles of the type used in the field experiments, viewed from above (Camera 2). Particles are generated at the nozzle at the lower left (green arrow) and are convected toward the right of the image.}}
\label{fig:rawSnow}
\end{figure}

{3D-PTV was used to quantify the movements of the snow particles. A wand-based calibration procedure, identical to that used in the field experiments (given in appendix \ref{sec:appB_calibration}), was used to reconstruct particle tracks in 3D. A wand composed of a pair of LEDs spaced 10 cm apart was moved throughout the measurement volume, and the calibration utility of \cite{theriault_protocol_2014} established the camera positions from image coordinates of the two lights in each camera view. The wand length was reconstructed with an error of 0.75\%, suggesting that the calibration was sufficiently accurate. The global coordinate system was set using a plumb line at the center of the wind tunnel, so that the nozzle was located at $(X,Y,Z) = (0, -0.2, 0)$ m. Particles were identified, triangulated in physical space, and tracked using the procedure outlined in appendix \ref{sec:appB_PTV}. Because of the high seeding density of the particles, only the largest 10\% by area of the identified particles in the raw images were triangulated into physical coordinates. This was done to reduce triangulation ambiguities, and to isolate the large particles that would have been observable in the field experiments for analysis. A statistical analysis of the size distributions of the particles from the two output settings was achieved by binarizing images taken by the camera positioned directly above the measurement volume (figure \ref{fig:rawSnow}), and converting the pixel areas of the identified particles to effective diameters in physical dimensions. The largest 10\% of the particles by area had effective diameters of $d_p = 11.2\pm4.2$ mm in the large-snow case and $d_p = 5.9\pm1.1$ mm in the small-snow case. For each experiment, at least 120 seconds of data were recorded, corresponding to over 30,000 raw images per camera. The resulting velocity vectors were binned and time-averaged into 3-cm cubic voxels with at least 10 vectors per displayed voxel.}

\subsection{Experimental Results}
\label{sec:appA_results}

\begin{figure}
\centering
\includegraphics[width=\textwidth]{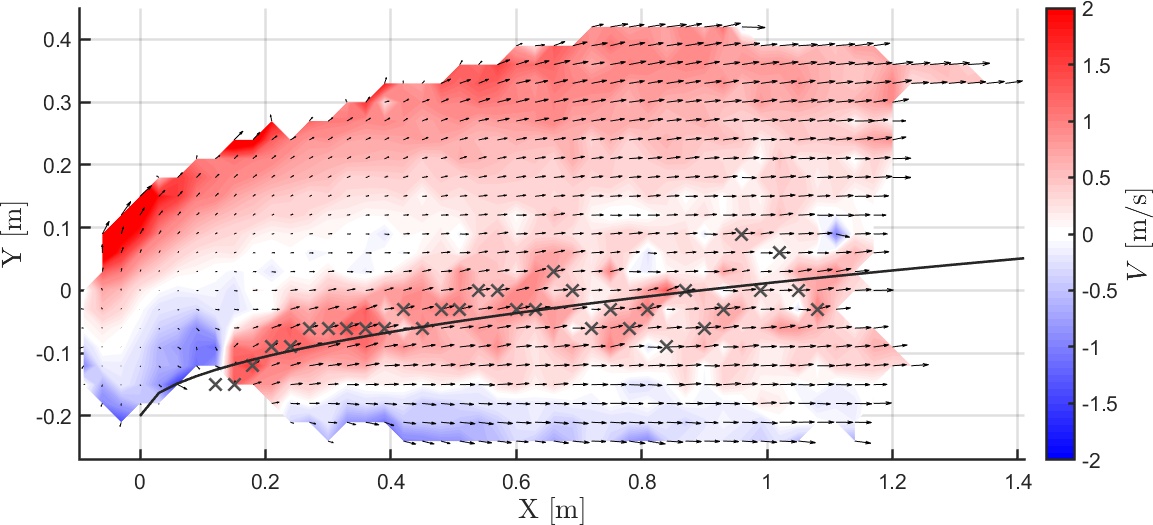}
\caption{{Contours of spanwise velocity $V$ at the plane $Z = 0$, for experiments with small snow particles at $U_\infty = 6.58\pm0.45$ $\rm{ms^{-1}}$. Grey crosses represent the identified crossflow-jet centerline from the data, and the black curve shows the resulting fit according to the profile given by \cite{hasselbrink_transverse_2001}. Negative velocities upstream of $X\lesssim0.1$ m were likely spurious, as particles in this region were clumped together and thus hard to identify accurately.}}
\label{fig:snowV}
\end{figure}

{Particle timescales and response characteristics were computed from the data in the following manner. First, the jet in crossflow was identified from velocity fields measured with the small snow particles. This was accomplished by applying a high-pass filter to streamwise slices of the spanwise velocity $V$ to isolate the jet in crossflow from the signal of the larger particle jet. A hill-climbing search algorithm was then applied within this region to find the local maximum of spanwise velocity corresponding to the center of the jet at each streamwise location. These maxima were projected into the $XY$ plane to define the jet centerline (figure \ref{fig:snowV}). To educe a jet profile from these data, it was assumed that the trajectory of the jet in crossflow would follow the similarity solution given by \cite{hasselbrink_transverse_2001} for the region near the jet orifice:} 

\begin{equation}
    {\frac{x}{rd}=\left(\frac{2}{c_{ej}}\frac{y}{rd}\right)^{1/2}.}
    \label{eqn:jet}
\end{equation}

{The jet diameter $d$ was estimated as 3 cm based on the diameter of the snow-machine particle generator, and the velocity ratio $r = \frac{V_{jet}}{U_\infty}$ was calculated based on the maximum velocity measured on the jet centerline. A single-parameter fit for the measured jet centerline was then applied to estimate the entrainment coefficient, $c_{ej}$. Between the two small-snow experiments ($r = 0.29$ and 0.36), the extracted entrainment coefficient was $c_{ej} = 0.39\pm0.01$. This was slightly larger than the value of $c_{ej}=0.32$ given by \cite{ricou_measurements_1961} for a free jet in controlled conditions, likely due to the higher turbulence intensities in this experiment. For the sake of consistency across experiments, this measured value for $c_{ej}$ was used to infer the trajectory of the jet in crossflow in the large-snow experiments, which had approximately the same jet diameter but lower velocity ratios as a result of the lower snow-machine blower setting.}

\begin{figure}
\centering
\includegraphics[width=\textwidth]{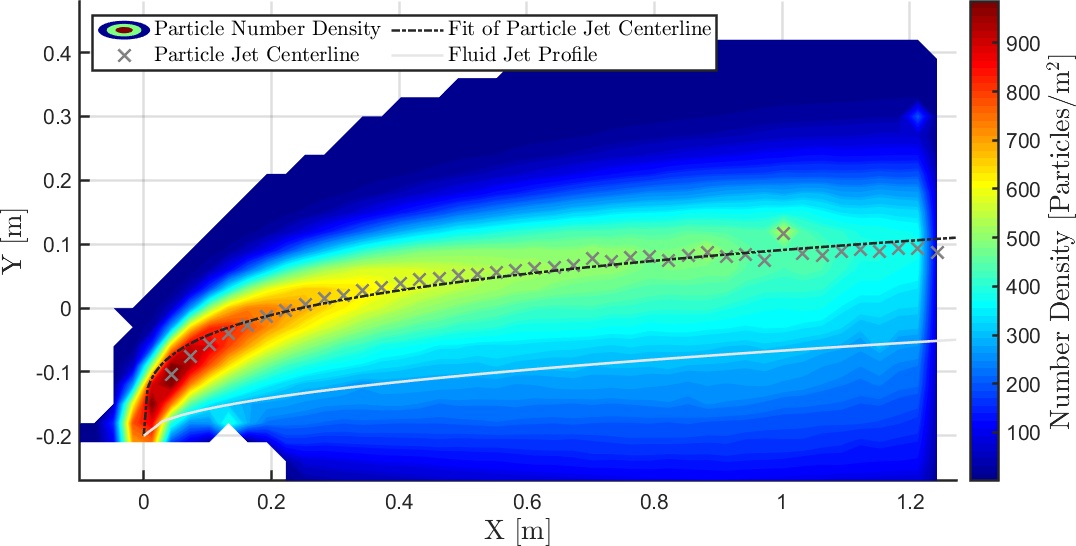}
\caption{{Contours of 2D particle number density at the plane $Z = 0$, for experiments with large snow particles at $U_\infty = 6.58\pm0.45$ $\rm{ms^{-1}}$. Grey crosses represent the identified particle-jet centerline from the data, and the black curve shows the resulting power-law fit, $y(x) = 0.291 x^{0.268} -0.2$. The profile of the jet in crossflow ($c_ej = 0.39$, $r = 0.116$) is given in light grey.}}
\label{fig:snowN}
\end{figure}

{The particle jet was identified using images taken from above the measurement domain (figure \ref{fig:snowN}). The particles detected in each frame were binned into 3-cm square areas and were plotted as a 2D distribution of particles, averaged over all images in the experiment. The center of the jet was identified at each streamwise location to subpixel accuracy using a three-point parabolic fit. The results were then fitted with a two-parameter power-law fit of the form $y(x) = ax^n - 0.2$, where another parabolic fit was applied along the line $Y = -0.2$ m to fix the coordinate system with respect to the jet orifice.}

{Given trajectories $y(x)$ for the crossflow and particle jets and information about the convective velocities along the trajectories, time-dependent profiles $x(t)$ and $y(t)$ could be inferred for the two jets. For the jet in crossflow, it was assumed that ideal tracer particles would follow the velocity profile derived from the aforementioned similarity solution of \cite{hasselbrink_transverse_2001} as}

\begin{equation}
    {U(x) = U_\infty \left(1-\frac{c_{vj}}{c_{ej}}\frac{d}{x}\right),}
    \label{eqn:jetU}
\end{equation}

\noindent{where the profile coefficient $c_{vj}$ was taken to be unity. A time history could then be extracted by integrating across point values of $\Delta t = \Delta x / U(x)$. For the particle jet, the streamwise velocity averaged across $Y$ and $Z$ at each streamwise location $X$ was taken to represent the average convective velocity of the particles at that location. This velocity followed a nonlinear relaxation in $X$ that was used to extract an average particle time history. The streamwise coordinates of the crossflow and particle jets, $x_{cf}(t)$ and $x_p(t)$, were computed for a common series of time steps by interpolation. Then, the spanwise coordinates $y_{cf}(t)$ and $y_p(t)$ were computed from $x(t)$ using the corresponding fit function for each jet profile.}

\begin{figure}
\centering
\includegraphics[width=0.48\textwidth]{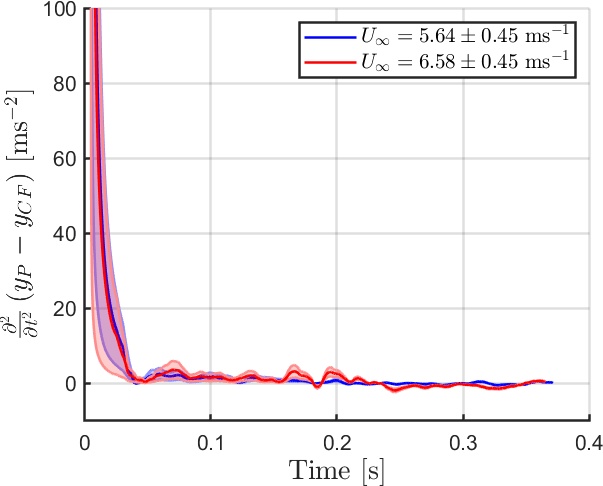}
\caption{{Impulse response in acceleration for the artificial snow particles used in the field experiments, at two free-stream velocities. Oscillations are artifacts of numerical errors from interpolation.}}
\label{fig:response}
\end{figure}

{From the matched time histories of $y_{cf}(t)$ and $y_p(t)$, a spanwise error was defined as $\delta(t) = y_p(t) - y_{cf}(t)$. This difference represents the deviation of an inertial particle from the background flow. Differentiating once with respect to time yielded the spanwise slip velocity of a particle subjected to a step change in $V$. Differentiating again in time gave the particle's impulse response in acceleration (figure \ref{fig:response}). The reliance on numerical differentiation required smooth signals to obtain meaningful results; it is for this reason that fit functions were used instead of measured data to obtain $y(t)$. In addition, a modest smoothing was required on $x(t)$ to prevent interpolation errors from accumulating through numerical differentiation. Error bars were computed from the residuals of the fits for $y_p(x)$ and $y_{cf}(x)$. The steady-state value of the profiles was computed as the mean of the latter half of the time history of the signal. The particle-response timescale $\tau_p$ was defined as the time at which the profiles of $\frac{\partial^2\delta}{\partial t^2}$ decayed to within one standard deviation of the converged value. The steady-state values of $\frac{\partial \delta}{\partial t}$ provided an estimation of the slip velocity $V_s$ of the particles. Computed particle timescales and slip velocities are given for the four experimental cases in table \ref{tab:snowResults}.}

\begin{table}
  \begin{center}
\def~{\hphantom{0}}
  \begin{tabular}{l|cccc}
        \textbf{Experimental Case} & {Small Snow 1} & {Small Snow 2} & {Large Snow 1} & {Large Snow 2} \\[6pt]
        Particle Diameter, $d_p$ (mm) & $5.9\pm1.1$ & $5.9\pm1.1$ & $11.2\pm4.2$ & $11.2\pm4.2$ \\
        Tunnel Speed, $U_\infty$ ($\rm{ms^{-1}}$) & $5.64\pm0.45$ & $6.58\pm0.45$ & $5.64\pm0.45$ & $6.58\pm0.45$ \\
        Particle Timescale, $\tau_p$ (ms) & $35.5\pm1.2$ & $34.4\pm2.0$ & $39.2\pm0.3$ & $40.5\pm6.7$ \\
        Slip Velocity, $V_s$ ($\rm{ms^{-1}}$) & $0.019\pm0.013$ & $0.025\pm0.017$ & $0.088\pm0.004$ & $0.101\pm0.030$ \\
  \end{tabular}
  \caption{{Particle diameters, timescales, and estimated slip velocities for the four cases in this experiment. Uncertainties represent one standard deviation from the mean quantities.}}
  \label{tab:snowResults}
  \end{center}
\end{table}

\subsection{Implications for Field Experiments}

{The small snow particles had slightly shorter timescales and smaller slip velocities than the large snow particles. For both types of particles, the particle-response timescales showed little dependence on the free-stream velocity, while the slip velocities increased modestly with $U_\infty$, most noticeably for the large snow particles. This was a consequence of the functional difference between the impulse response in acceleration, which does not depend on the magnitude of the disturbance, and the step response in velocity, which does. The particle-response timescales for the large snow particles were therefore representative of those expected in the field experiments. The relevant flow timescale in the field experiments was $\tau_f = D/U_\infty\approx0.18$, resulting in a particle Stokes number of $Sk \approx 0.23$. Since $\tau_p < \tau_f$, it could be assumed that the large snow particles used in the field experiments would respond rapidly enough to resolve the flow structures of interest to this study.}

{Since the slip velocities showed a tendency to increase with the free-stream velocity, the particle slip velocities in the field experiments were expected to be somewhat larger than those measured in the laboratory experiments. Assuming the particle slip velocity is directly proportional to $U_\infty$, a linear extrapolation to the average wind speed at the field site ($U_\infty = 11$ $\rm{ms^{-1}}$) yields a worst-case particle slip velocity of $V_{s,max} = 0.170$ $\rm{ms^{-1}}$. The acceleration-impulse condition produced in the wind-tunnel crossflow configuration was somewhat dramatic compared to conditions encountered by the particles in the field experiments, in which the jet was aligned with the free-stream flow. Therefore, this slip velocity should be treated as an upper bound on the flow fidelity of the particles. Errors due to particle slip in the field experiments were thus estimated to be below 2\% of the free-stream velocity, which was similar to the precision of the 3D-PTV measurement system (cf.\ appendix \ref{sec:appB_PTV}).}

{Lastly, the effect of the snow machines on the flow incident on the wind turbine in the field experiment can be ascertained from the results of the laboratory experiments. The maximum velocity of the flow exiting the snow machine for the large-snow case was $0.993\pm0.165$ $\rm{ms^{-1}}$. Since the snow machines expelled their particles in the streamwise direction in the field experiments, this injection of momentum was minimal compared to the average wind speed. In addition, the particle-response timescales measured in the laboratory suggest that the particles would have equilibrated to the background flow well in advance of their approach to the turbine (on the order of 1 m from the snow-machine nozzle). Given the turbulent nature of the wind conditions at the field site, the presence of the snow machines was not expected to disrupt the inflow condition to the turbine. These considerations suggested that the effect of the snow machines on the quality and measurement of the flow conditions was negligible.}

\section{Processing Procedures}
\label{sec:appendix_processing}

\subsection{Camera Calibration}
\label{sec:appB_calibration}

Calibrations were performed before each turbine was raised into position to facilitate wand motion throughout the measurement volume. One calibration was used for all measurements with the UGE turbine, and a second calibration was used for all measurements with the WPE turbine. A wand, consisting of two 2,000-lumen LEDs (XLamp XM-L, Cree Components) spaced 1.15 m apart, was moved throughout the measurement volume using a reach forklift. A light located in the middle of the wand was flashed to provide a synchronization signal for the cameras. Using the MATLAB tool developed by \cite{theriault_protocol_2014}, the LED positions and wand lengths were reconstructed in an arbitrarily assigned global coordinate system. The standard deviations of the reconstructed wand lengths from the two calibrations, over all recorded images of the wand as it moved through the measurement domain, were $0.16\%$ and $0.20\%$ of the measured wand length. The accuracy of the calibrations was further examined by comparing the calculated and measured distances between cameras, to quantify reprojection errors. The average reprojection errors across all inter-camera distances were $0.74\pm0.39\%$ and $0.83\pm0.41\%$ for the two calibrations. These corresponded to less than 10 cm in physical space, which was within the error tolerances of the tape-based physical measurements themselves. Lastly, the two calibrations were compared by applying both calibrations to the videos of the first calibration. After matching the two coordinate systems using principal-component analysis, the average distance between corresponding LED positions in the two calibrations was 1.34 cm. This corresponded to a difference on the order of 1 pixel in the camera images, suggesting that the difference between the two calibrations was negligible. Still, for the sake of consistency, the first calibration was used for all of the experiments with the UGE turbine, while the second calibration was used for all experiments with the WPE turbine. As a final check on the validity of the calibrations, images in which the turbines were present in the camera views were used to reconstruct the spans of the turbines. Several points along the top and bottom of each turbine were selected in several camera views, and an elliptical fit of these points identified points lying on the central axis of the turbine. Triangulating the positions of these points in 3D space (using the methods outlined later in section \ref{sec:sec2_analysis}) yielded turbine spans that differed by $0.21\%$ and $0.32\%$, respectively, from the actual values. Overall, these analyses demonstrated that both calibrations were accurate and consistent in their 3D reconstruction of objects in the measurement domain.

\subsection{3D-PTV Processing and Statistical Convergence}
\label{sec:appB_PTV}

To obtain accurate time-averaged velocity fields from the raw camera images, a series of processing steps were undertaken. First, for every individual data set, the images from each camera were averaged to produce a background image, which was then subtracted from each image to remove glare and stationary structures. Then, the turbine and support structures visible in each frame were masked using an intensity threshold and object-detection routine. Through this combination of background subtraction and masking, the {artificial snow} particles in each image were isolated. The results of these steps were checked manually for each data set to confirm that the correct regions of the flow field were isolated.

{To compute the temporal offset between images from different cameras, the signal of the synchronization light on the turbine tower was tracked in each image.} The mean of the pixel intensities in a small region containing the synchronization light was computed for every frame in each set of camera images. A custom edge-finding routine was then used to locate the frames in each camera view in which the light was switched on. The six sets of images were temporally aligned according to these reference frames, and the subset of time instances that were represented in all six camera views was selected for the particle-identification step.

To identify particles in this set of images, intensity thresholds were chosen for every camera in each data set. These thresholds were used to binarize the images, and were set manually by iterating on a subset of the images in a data set, so that approximately 200 particle candidates would be identified in each image. This target number of particles was found to be sufficiently high to remain sensitive to poorly illuminated particles, while low enough to prevent false positives from obfuscating actual particles. Pixel coordinates for each particle candidate were determined by computing its centroid, which was more robust to the non-spherical nature of the particles than other measures. These identified coordinates in 2D image space were then triangulated onto 3D global coordinates based on the calibrations outlined in section \ref{sec:sec2_PTV}, using epipolar geometry \citep{hartley_multiple_2003}. To reduce the number of ghost particles (nonphysical artifacts from ambiguities in epipolar geometry) detected by this triangulation approach, it was required that a particle appear in at least three camera views in order for it to be triangulated into 3D space \citep{elsinga_velocity_2010}. Particle trajectories and velocities were then numerically computed from these spatial coordinates for all time instances in the data set using a multi-frame predictive-tracking algorithm developed by \cite{ouellette_quantitative_2006} and \cite{xu_tracking_2008}. The velocity fields for three data sets with different snow-machine heights were then combined into a single unstructured 3D volume of three-component velocity vectors. This volume was rotated into the coordinate system given in figure \ref{fig:fieldSite} from the arbitrary one assigned by the calibrations using the axis of the turbine and the tips of the snow-machine towers as references. The result of this procedure was a collection of velocity vectors distributed throughout the measurement domain, representing instantaneous velocity measurements at various time instances during the experiment.

\begin{figure}
\begin{subfigure}[t]{0.48\textwidth}
\centering
  \includegraphics[width=\textwidth]{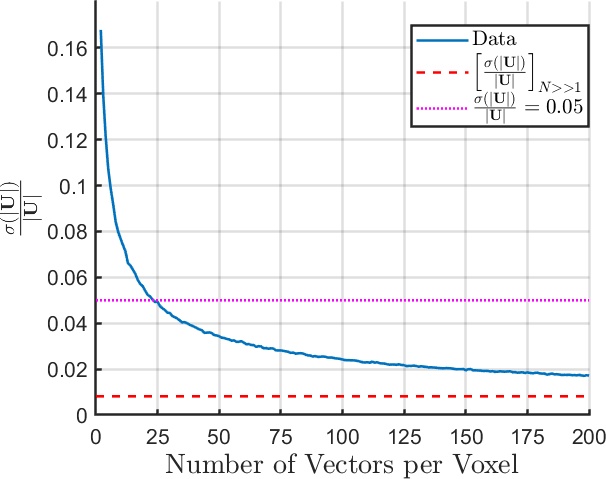}
  \caption{}
\label{fig:stdMean}
\end{subfigure}
\hfill
\begin{subfigure}[t]{0.48\textwidth}
\centering
  \includegraphics[width=\textwidth]{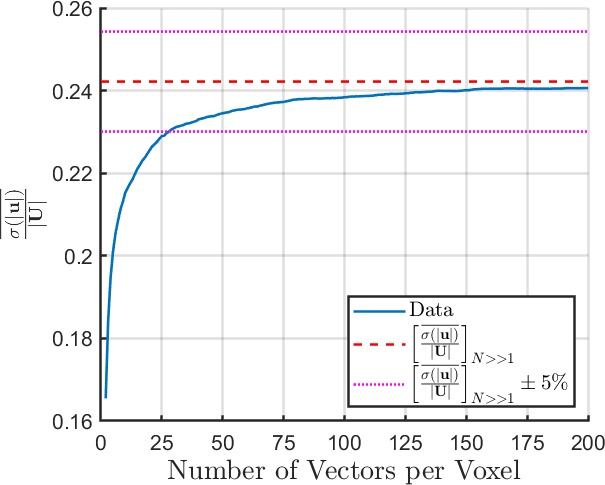}
  \caption{}
\label{fig:meanStd}
\end{subfigure}
\caption{Statistical analysis of vectors contained in a 25-cm cubic voxel, located 1.5 $D$ upstream of the UGE turbine. The variations of the (a) standard deviation of bootstrapped means and (b) mean of bootstrapped standard deviations of the velocity magnitude are shown against the number of vectors taken in each sample. In both figures, the converged value of each measure for $N>>1$ is shown as a red dashed line, while bounds for acceptable convergence are given as dotted magenta lines. {$N \gtrsim 25$ yields convergence within 5\%, while $N \gtrsim 150$ yields convergence within 2\%.}}
\label{fig:stats25cm}
\end{figure}

The unstructured velocity vectors were then interpolated onto a grid of cubic voxels to produce a single time-averaged velocity field. As the choice of the voxel size influenced both the resolution and statistical convergence of the measured velocity field, statistical analyses were carried out to inform this selection. First, the effect of the number of vectors per voxel on statistical convergence was ascertained. For this analysis, the first case of the helical-bladed UGE turbine was selected. Bootstrap sampling was employed to obtain better estimates of the statistics of the population \citep{efron_bootstrap_1979}. For a voxel with a side length of 25 cm located upstream of the turbine, 9,000 bootstrap samples of $N$ vectors were taken, and the mean and standard deviation of the velocity magnitude were computed for each sample. This process was repeated for values of $N$ from 1 to the total number of vectors in that voxel. The standard deviation of these bootstrapped means represented the uncertainty due to computing an average from samples of the entire population. This uncertainty decreased with increasing $N$, and dropped below $5\%$ of the average velocity magnitude of all samples in the voxel, $|\mathbf{U}|$, at $N \approx 25$ (figure \ref{fig:stdMean}). {The uncertainty fell below $2\%$ at $N \approx 150$.} A similar conclusion regarding the value of $N$ was drawn from the average of the bootstrapped standard deviations for various values of $N$ (figure \ref{fig:meanStd}). Hence, subsequent analyses sought to include 25 vectors per voxel, where possible.



\begin{figure}
\begin{subfigure}[t]{0.48\textwidth}
\centering
  \includegraphics[width=\textwidth]{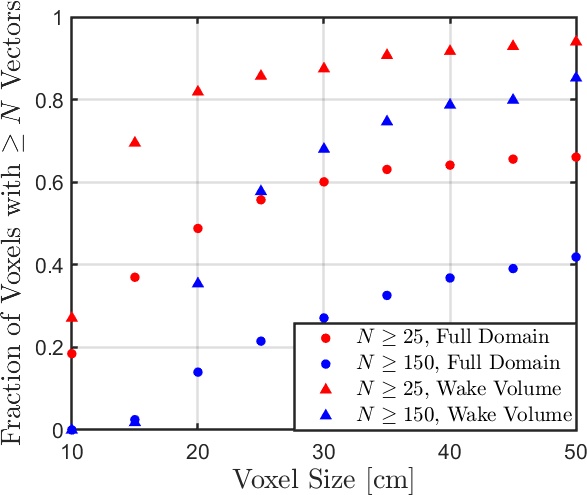}
  \caption{}
\label{fig:vecPerVox}
\end{subfigure}
\hfill
\begin{subfigure}[t]{0.48\textwidth}
\centering
  \includegraphics[width=\textwidth]{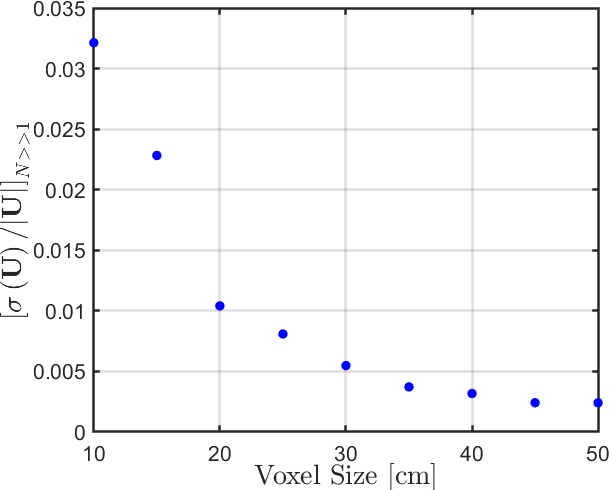}
  \caption{}
\label{fig:voxMeans}
\end{subfigure}
\caption{Measures for the determination of an appropriate voxel size for binning and averaging velocity vectors. (a) shows the fraction of voxels containing at least $N = 25$ and $N = 150$ vectors, {representing the number of vectors required for 5\% and 2\% measurement precision, for the entire measurement domain (circles) and the wake region (triangles).} (b) shows the standard deviation of bootstrapped means for all vectors in a given voxel, {representing the best-case precision possible for a given voxel size}. Both measures suggest that a grid dimension of 25 cm is a good compromise between spatial resolution and statistical convergence.}
\label{fig:statsVox}
\end{figure}

Given that the distribution of particle traces in the domain was not uniform, a series of voxel sizes was tested on the same data set used above to determine a voxel size that balanced spatial resolution and the target number of vectors per voxel. {Voxels of various sizes were used to discretize the domain, and voxels that contained at least 3 vectors were counted toward the total number of voxels in each discretization.} The fraction of {these voxels that contained} at least 25 vectors began to converge around a grid dimension of 30 cm (figure \ref{fig:vecPerVox}). {A grid dimension of 25 cm resulted in over 50\% of voxels having at least 25 vectors (5\% precision, according to figure \ref{fig:stdMean}), as well as over 20\% having at least 150 vectors (2\% precision). In a volume encapsulating the turbine wake (bounded by $X/D > \frac{1}{2}$, $\left|Y/D\right| \leq \frac{3}{2}$, and $\left|Z/D\right| \leq \frac{3}{2}$), 87\% of voxels had at least 25 vectors, and 58\% of voxels had at least 150 vectors. Therefore, the level of precision of the results presented in this work, which focus on this wake volume, was comfortably below 5\% in the area of interest for voxels of this size.} In addition, the standard deviation of the bootstrapped means for \textit{all} vectors within a voxel of a given size dropped below $1\%$ of $|\mathbf{U}|$ for a grid dimension of 25 cm (figure \ref{fig:voxMeans}). This indicated very good {best-case} statistical convergence for voxels of at least this size. 

{In light of these results, a voxel size of 25 cm was used to discretize the domain, serving as a good compromise between grid resolution and statistical convergence. At least half of all voxels in the measurement volume had standard deviations in the velocity magnitude below 5\%, with 20\% of these having a measurement precision below 2\%. In the wake of the turbine, the precision for 87\% of the voxels was below 5\%, with close to 60\% of this volume having a precision below 2\%. The best-case precision for the most densely populated voxels was below 1\%.}


\begin{figure}
	\begin{subfigure}[t]{0.45\textwidth}
\centering
  \includegraphics[width=\textwidth]{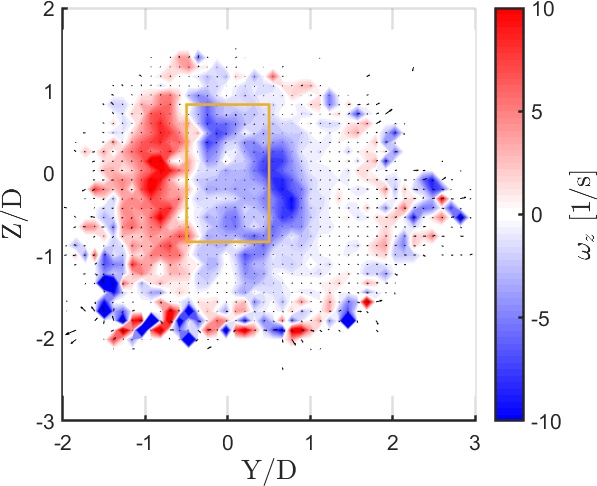}
  \caption{No filtering}
\label{fig:nofilter}
\end{subfigure}
\hfill
\begin{subfigure}[t]{0.45\textwidth}
\centering
  \includegraphics[width=\textwidth]{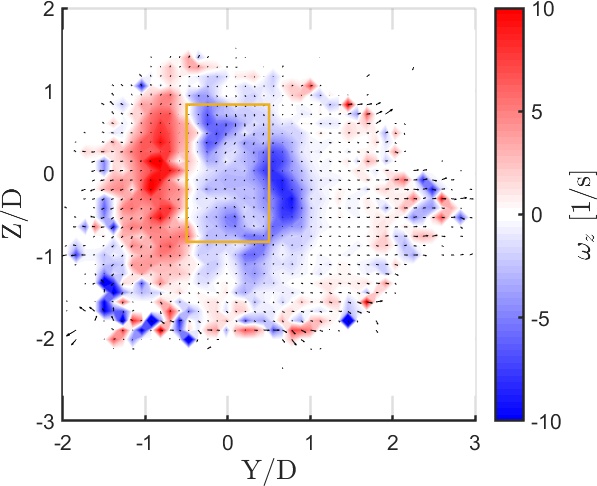}
  \caption{Solenoidal filtering only}
\label{fig:solfilter}
\end{subfigure}
\begin{subfigure}[t]{0.45\textwidth}
\centering
  \includegraphics[width=\textwidth]{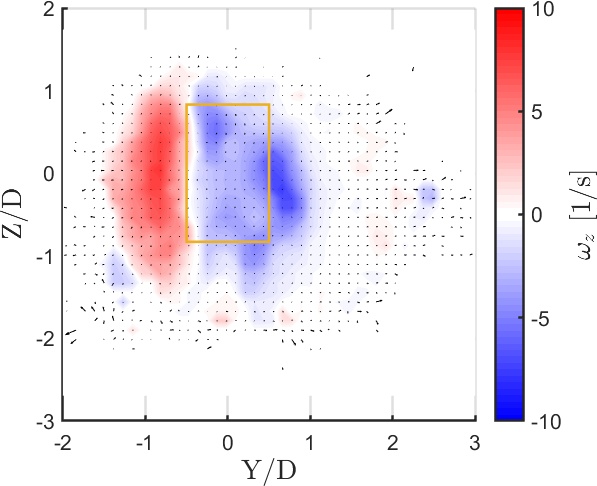}
  \caption{Median filtering only}
\label{fig:medfilter}
\end{subfigure}
\hfill
\begin{subfigure}[t]{0.45\textwidth}
\centering
  \includegraphics[width=\textwidth]{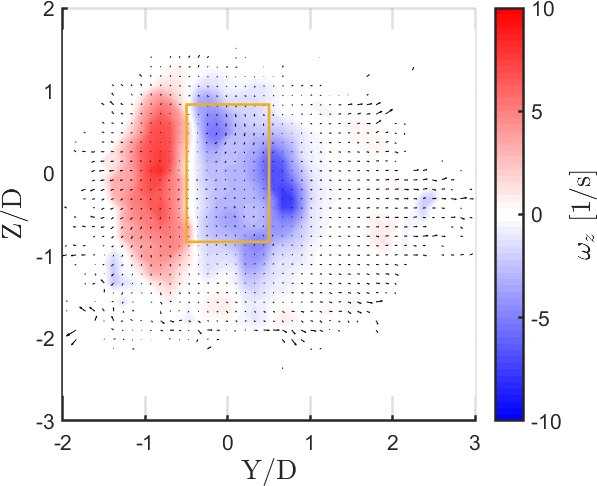}
  \caption{Both solenoidal and median filtering}
\label{fig:allfilter}
\end{subfigure}
    \caption{Effects of the two filters applied to the voxel-averaged velocity and vorticity fields, demonstrated on a cross-section of vertical vorticity ($\omega_z$) at $X/D = 1.5$ downstream of the WPE turbine for $\lambda = 1.20$. The solenoidal filter affects both the velocity and vorticity readings, while the median filter is only applied to the vorticity field.}
    \label{fig:filters}
\end{figure}

Once the velocity vectors had been binned and averaged into 25-cm cubic voxels, a filter developed by \cite{schiavazzi_matching_2014} was employed to enforce a zero-divergence criterion on the vector fields, in keeping with the negligible Mach numbers ($Ma < 0.05$) of the experimental conditions. The effect of this filter was to attenuate unphysical deviations in the velocity fields due to spurious particle trajectories, especially near the edges of the domain where velocity vectors were more sparse (cf.\ figure \ref{fig:filters}).

From these filtered velocity fields, values for the velocity incident on the turbine and the settling velocity of the artificial snow particles were computed by averaging all velocity vectors at least 1 $D$ upstream of the axis of rotation of the turbine. Compared to the settling velocity of the particles measured in quiescent air, $W_s^0 = 0.60\pm0.18$ $\rm{ms^{-1}}$, the observed settling velocities measured in the prevailing wind conditions at the field site were higher: $W_s = 0.89\pm0.12$ $\rm{ms^{-1}}$. The discrepancy was likely due to the presence of a slight downward slope in the local topography at the site. According to data collected by \cite{kinzel_energy_2012}, the corresponding bias in the vertical velocity was $W \approx -0.22$ $\rm{ms^{-1}}$, which is consistent with the discrepancy observed in the present data. {It is also possible that atmospheric turbulence contributed to the increased settling velocity as well \citep{nemes_snowflakes_2017}.} Thus, the average settling velocity measured upstream of the turbine, $W_s$, was subtracted from the entire velocity field, so that the time-averaged vertical velocity $W$ induced by the turbine could be isolated.

Fields of vorticity were computed from these velocity fields. Due to the relatively coarse grid size and the error associated with numerical differentiation, a $3\times3\times3$ median filter was applied to all fields involving velocity derivatives. This was mainly employed to remove unphysical results from numerical differentiation near the edges of the measurement domain, but also had a modest smoothing effect on the vortical structures present in the wake. 

To demonstrate the effects of the filters applied to the velocity and vorticity fields, a single planar slice at $X/D = 1.5$ was isolated from the WPE turbine data at $\lambda = 1.20$, and the vertical vorticity component $\omega_z$ was plotted in figure \ref{fig:filters}. The unfiltered vorticity fields were marked by significant noise on the boundaries of the domain (figure \ref{fig:nofilter}). Applying the solenoidal filter to the velocity fields resulted in locally smoothed velocity vectors, corresponding to very slight smoothing in the vorticity fields (figure \ref{fig:solfilter}). By contrast, applying the $3\times3\times3$ median filter to the vorticity fields removed the shot noise from the boundaries of the domain, while smoothing over the large-scale structures (figure \ref{fig:medfilter}). The combination of the two filters resulted in smoothed velocity and vorticity fields that allowed trends in the flow fields to be identified more readily (figure \ref{fig:allfilter}).

\section{Velocity and Vorticity Fields}
\label{sec:appendix_fields}

In this section, salient features of the velocity and vorticity fields in the wake are described and discussed. These results are compared with previous studies in the literature to show that many of the trends observed in laboratory experiments at lower Reynolds numbers still apply for full-scale turbines in field conditions.

\subsection{Velocity Fields}
\label{sec:appC_velocity}

\begin{figure}
\begin{subfigure}[t]{0.48\textwidth}
\centering
  \includegraphics[width=\textwidth]{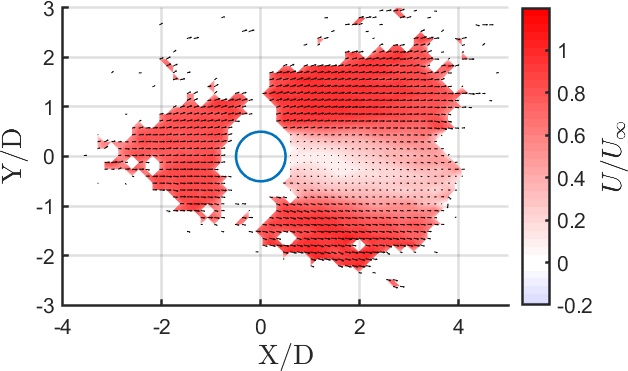}
  \caption{}
\label{fig:u_UGE1}
\end{subfigure}
\hfill
\begin{subfigure}[t]{0.48\textwidth}
\centering
  \includegraphics[width=\textwidth]{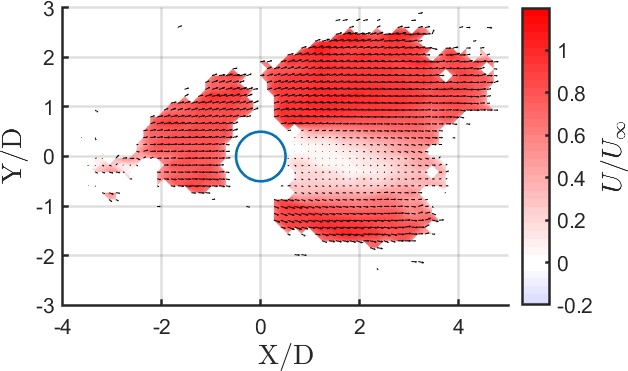}
  \caption{}
\label{fig:u_UGE2}
\end{subfigure}
\caption{Time-averaged planar fields of the streamwise velocity $U$ for the UGE turbine at (a) $\lambda = 1.19$ and (b) $\lambda = 1.40$, taken at $Z/D=0$. The differences in the shape of the wake between the two tip-speed ratios are minor.}
\label{fig:u_UGE_XY}
\end{figure}

\begin{figure}
\begin{subfigure}[t]{0.48\textwidth}
\centering
  \includegraphics[width=\textwidth]{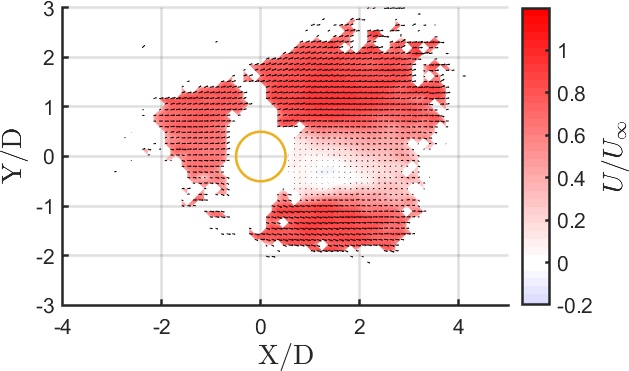}
  \caption{}
\label{fig:u_WPE1}
\end{subfigure}
\hfill
\begin{subfigure}[t]{0.48\textwidth}
\centering
  \includegraphics[width=\textwidth]{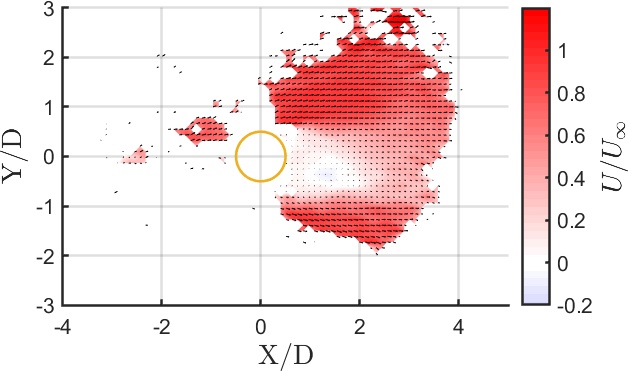}
  \caption{}
\label{fig:u_WPE2}
\end{subfigure}
\caption{Time-averaged planar fields of the streamwise velocity $U$ for the WPE turbine at (a) $\lambda = 0.96$ and (b) $\lambda = 1.20$, taken at $Z/D=0$. As in figure \ref{fig:u_UGE_XY}, the differences in the wake between these two tip-speed ratios are minor.}
\label{fig:u_WPE_XY}
\end{figure}

Velocity fields for the time-averaged streamwise-velocity component $U$ on three orthogonal planar cross-sections are given for the UGE and WPE turbines respectively, each at two tip-speed ratios, in figures \ref{fig:u_UGE_XY} and \ref{fig:u_WPE_XY}. The velocity deficit in the wakes were slightly more pronounced in the higher-$\lambda$ cases, and the wake regions were shifted slightly toward the negative spanwise direction due to the rotation of the turbine. This is consistent with the wake trends observed by \cite{parker_effect_2016}. The magnitude of the velocity deficit was significantly larger for the WPE turbine, consistent with its higher solidity. These flow fields qualitatively paralleled those reported by \cite{araya_transition_2017}, which were recorded in a water channel at a significantly lower Reynolds number ($Re_D = 8\times10^4$). This supports the inference of \cite{parker_effect_2016} that the general shape of the turbine wake is not very sensitive to Reynolds number in this parameter range, despite the fact that the coefficients of power only converge for $Re_D \gtrsim 1.5\times10^6$ \citep{miller_vertical-axis_2018}.

\begin{figure}
\begin{subfigure}[t]{0.48\textwidth}
\centering
  \includegraphics[width=\textwidth]{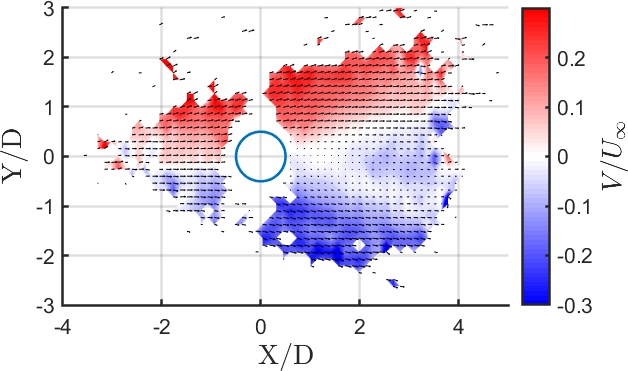}
  \caption{}
\label{fig:v_UGE}
\end{subfigure}
\hfill
\begin{subfigure}[t]{0.48\textwidth}
\centering
  \includegraphics[width=\textwidth]{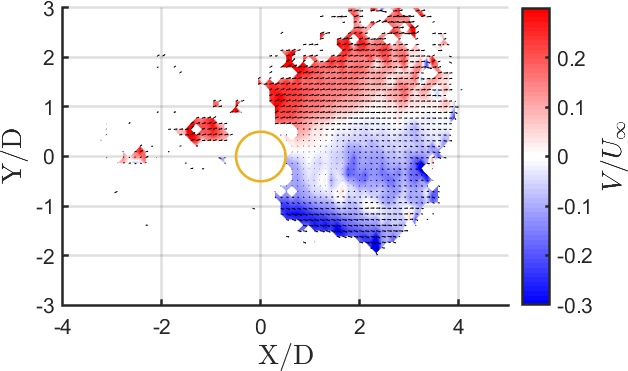}
  \caption{}
\label{fig:v_WPE}
\end{subfigure}
\caption{Time-averaged planar fields of the spanwise velocity $V$ for (a) the UGE turbine at $\lambda = 1.19$ and (b) the WPE turbine at $\lambda = 1.20$, taken at $Z/D=0$. The V-shaped region of negative spanwise velocity downstream of the turbines is more prominent for the WPE turbine, which has a higher solidity.}
\label{fig:v}
\end{figure}

Velocity fields for the time-averaged spanwise-velocity component $V$, also taken at the mid-span of the turbines, are shown in figures \ref{fig:v_UGE} and \ref{fig:v_WPE} for the UGE and WPE turbines at $\lambda=1.2$. A clear upstream bifurcation of the flow due to the presence of the turbine was visible in figure \ref{fig:v_UGE}, and a negatively skewed velocity field, representing spanwise flow induced by the rotation of the turbine, was present downstream of the turbine in both cases. This region grew more rapidly in thickness behind the WPE turbine than behind the UGE turbine, again likely due to its higher solidity and correspondingly stronger flow induction.

\begin{figure}
\centering
\includegraphics[width=0.5\textwidth]{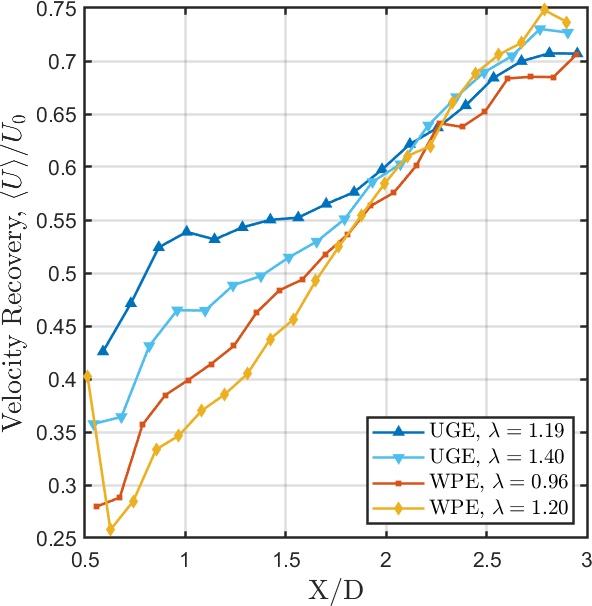}
\caption{Profiles of $\langle U \rangle / U_0$ versus distance downstream of the turbine for all four experimental cases. Here, angle brackets denote spatial averages across $YZ$-sections of the wake, and $U_0$ represents the velocity directly upstream of the turbine. Profile discrepancies corresponding to differences in $\lambda$ and $\sigma$ are present in the near wake ($X/D \lesssim 2$), whereas the wake recovery in the far wake appears to be more uniform.}
\label{fig:wakeRecovery}
\end{figure}

The wake recovery of the two turbines also followed previously observed trends in the literature \cite[cf.][]{ryan_three-dimensional_2016,araya_transition_2017}. The wake recovery was quantified by thresholding planar slices of the streamwise velocity component at various streamwise positions downstream of the turbine by the average upstream flow velocity incident on the turbine, $U_0$. For each slice, this thresholding procedure divided the wake region, defined as the region of flow where the local streamwise velocity component was less than $U_\infty$, from the surrounding free-stream region. The average velocity within the wake region, $\langle U\rangle$, was taken for each streamwise slice, and was plotted against streamwise distance (figure \ref{fig:wakeRecovery}). The analysis of these wake-recovery profiles has been undertaken comprehensively by \cite{araya_transition_2017}, and the trends found in these experiments show good agreement with their findings. The minimum value of $\langle U \rangle / U_0$ was observed to decrease with increasing $\lambda$ and increasing $\sigma$. These differences were most prominent in the near wake ($X/D \lesssim 2$), in which vortex shedding from the turbine blades is most significant \citep{tescione_near_2014,parker_effect_2016,araya_transition_2017}. The streamwise extent of the measurement domain was not large enough to observe the full transition to bluff-body wake dynamics described by \cite{araya_transition_2017}, but the presently measured wake profiles exhibit similarity to that previous work within the domain of present interest.

\subsection{Vorticity Fields}
\label{sec:appC_vorticity}

\begin{figure}
\centering
\includegraphics[width=\textwidth]{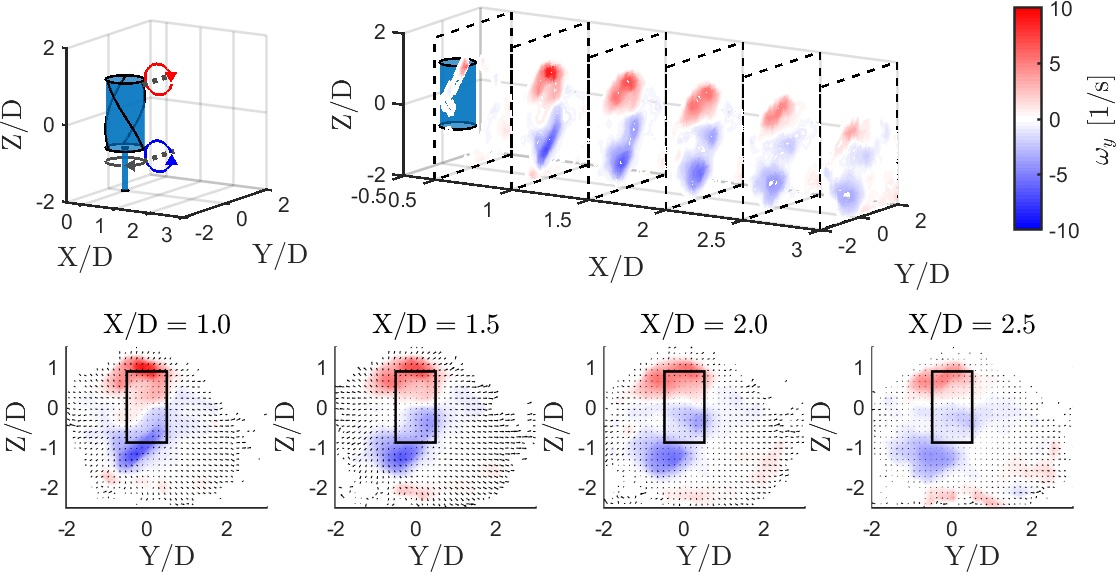}
\caption{Streamwise slices of the spanwise vorticity $\omega_y$ downstream of the UGE turbine for $\lambda = 1.19$. These structures are products of vortex shedding from the tips of the turbine blades \citep{tescione_near_2014}. Note that the $X$-axis is stretched on $0.5\leq X/D \leq 3$.}
\label{fig:wy_UGE}
\end{figure}

\begin{figure}
\centering
\includegraphics[width=\textwidth]{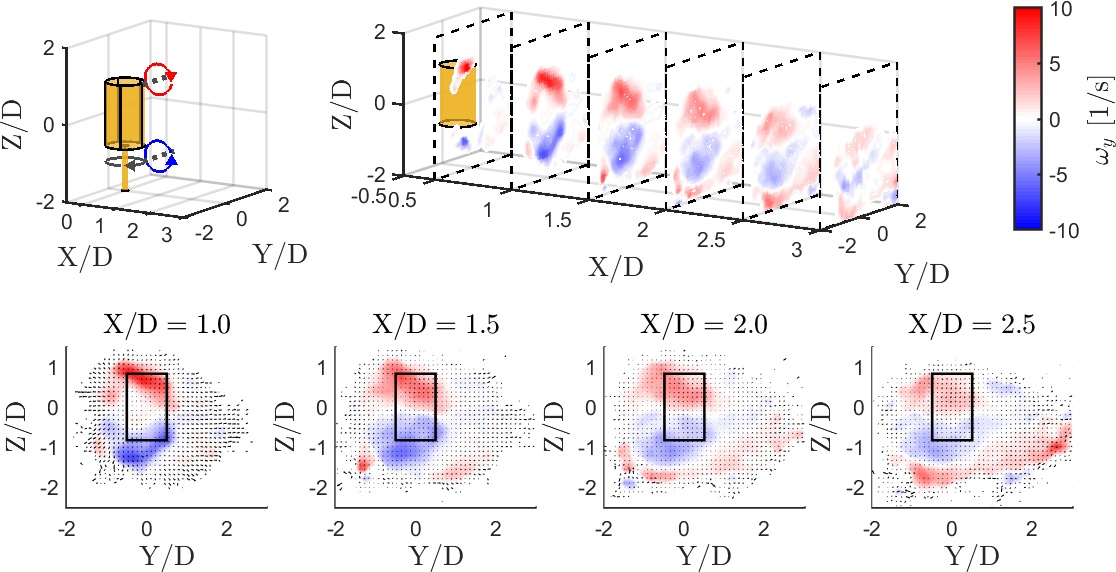}
\caption{Streamwise slices of the spanwise vorticity $\omega_y$ downstream of the WPE turbine for $\lambda = 1.20$. These structures are not significantly different from those shown in figure \ref{fig:wy_UGE}.}
\label{fig:wy_WPE}
\end{figure}

For the sake of completeness, streamwise slices of the spanwise vorticity $\omega_y$ are provided in figures \ref{fig:wy_UGE} and \ref{fig:wy_WPE}. The time-averaged vortical structures visible in these plots represented tip vortices shed from the turbine blades \citep{tescione_near_2014}. These structures did not display a strong degree of asymmetry, suggesting that they were relatively unaffected by the tilted-wake behavior observed in the vortical structures in $\omega_x$ and $\omega_z$.

\bibliographystyle{jfm}
\bibliography{Field_PTV_Paper}

\end{document}